\documentclass[useAMS,usenatbib]{mn2e}
\usepackage{ mathrsfs }
\usepackage{ gensymb }
\usepackage{graphicx}
\usepackage{epsfig}
\usepackage{ tipa }
\usepackage{threeparttable}
\usepackage{graphicx, subfigure}

\title[M dwarf activity as noise in exoplanet detection]
  {Stellar activity as noise in exoplanet detection II.  Application to M dwarfs
}
\author[J.M. Andersen \& H. Korhonen]
  {J.M.~Andersen,$^{1,2}$\thanks{Email: janmarie@bu.edu}
  H.~Korhonen,$^{3,4,2}$ \\
  $^1$Department of Astronomy, Boston University, 725 Commonwealth Avenue, Boston, MA 02215, USA\\
  $^2$Centre for Star and Planet Formation, Natural History Museum of Denmark, University of Copenhagen, \O ster Voldgade 5-7, \\DK-1350, Copenhagen, Denmark\\
  $^3$Finnish Centre for Astronomy with ESO (FINCA), University of Turku, V\"{a}is\"{a}l\"{a}ntie 20, FI-21500 Piikki\"{o}, Finland\\
  $^4$Niels Bohr Institute, University of Copenhagen, Juliane Maries Vej 30, DK-2100 Copenhagen, Denmark}

\date{Released 2014 Xxxxx XX}

\pagerange{\pageref{firstpage}--\pageref{lastpage}} \pubyear{2002}

\def\LaTeX{L\kern-.36em\raise.3ex\hbox{a}\kern-.15em
    T\kern-.1667em\lower.7ex\hbox{E}\kern-.125emX}
\begin{document}

\label{firstpage}

\maketitle

\begin{abstract}

The ubiquity of M dwarf stars combined with their low masses and luminosities make them prime targets in the search for nearby, habitable exoplanets. We investigate the effects of starspot-induced radial velocity (RV) jitter on detection and characterization of planets orbiting M dwarfs. We create surface spot configurations with both random spot coverage and active regions. Synthetic stellar spectra are calculated from a given spot map, and RV measurements are obtained using cross-correlation technique. We add the RV signal of an orbiting planet to these jitter measurements, and reduce the data to ``measure" the planetary parameters. We investigate the detectability of planets around M dwarfs of different activity levels, and the recovery of input planetary parameters. When studying the recovery of the planetary period we note that while our original orbital radius places the planet inside the HZ of its star, even at a filling factor of 2\% a few of our measurements fall outside the ``conservative Habitable Zone". Higher spot filling factors result in more and higher deviations. Our investigations suggest that caution should be used when characterizing planets discovered with the RV method around stars that are (or are potentially) active.
	  
\end{abstract}


\section{Introduction}

M dwarfs are of great interest in the search for nearby habitable planets for many reasons.  They have been hailed by some as the holy grail of stars for potential life-permitting-planet hosts, but have  been condemned by others as unsuitable targets for planet searches.  Due to their ubiquity alone--M dwarfs make up $\sim$70 percent of stars in the Milky Way \citep{Henry1994, Chabrier2003, Reid2004, Covey2008, Bochanski2010}, combined with the fact that observations indicate the occurrence rate of super-Earths and Earth-sized planets increases with decreasing stellar mass \citep{Howard2012}, the discovery of a nearby habitable planet seems most likely to be found orbiting an M dwarf. A recent study by \citet{Dressing2013} found a lower limit on the occurrence rate of Earth-size planets in the habitable zones (HZ) of cool stars of 0.04 planets per star, and also determined that the nearest planet in the habitable zone should be just 5 pc away, although \cite{Mann2013} contests that this estimate might be too high based on their revised values of $R_{*}$, $L_{*}$, $M_{*}$ for a sample of Kepler stars, which resulted in some of the formerly HZ planets no longer orbiting within the HZ.  

The HZ is most loosely defined as the area around a star in which liquid water could exist on the planet's surface.  According to the model developed by \cite{Kopparapu2013}, the inner and outer HZ limits for our Solar System are 0.99 AU and 1.70 AU, respectively.  \cite{Kopparapu2013} point out that their model does not include the radiative effects of clouds, so the boundaries could extend farther.  
 M dwarfs are much less massive and less luminous than solar-type stars, which means the HZ is significantly closer to the star (for example, for an M dwarf of mass  $M \approx 0.5M_{\odot}$, and $R \approx 0.5R_{\odot}$, the HZ is between $\sim$0.1 and $\sim$0.2\,AU). This makes small planets orbiting in the HZ around M dwarfs easier to detect than their counterparts orbiting more massive stars for a few reasons:
(1) Transits are more likely to be detected due to the the HZ being significantly closer to the star \citep{Kasting1993}, which results in a larger range of inclination angles that will yield a visible transit from Earth.  A HZ planet orbiting a $0.5M_{\odot}$,  $0.5R_{\odot}$ M dwarf has a $\sim$3 times higher likelihood of transiting compared to a planet orbiting a Solar--mass star in the star's HZ.  
(2) A larger planet--to--star ratio results in larger transit depths in the light curve, since the transit depth is simply proportional to the area of the star that is being eclipsed by the transiting planet. An Earth-mass planet in the HZ of a $R = 0.5R_{\odot}$ M dwarf would have $\sim$4 times the transit depth compared to a similar planet orbiting a Sun--like star.  
(3) The close--in habitable zone around M dwarfs also makes radial velocity (RV) reflex motion larger (and thus in principle easier to measure).  Planets of a given mass would be easier to detect around M dwarfs than around larger stars because of the reduced mass ratio of the system. An Earth-mass planet orbiting in the HZ of a $0.5M_{\odot}$ M dwarf would have a Doppler shift 3 to 4 times stronger than the same planet orbiting in the Sun's HZ.

However, a HZ near to the star can cause a host of other potential problems. For instance, a planet orbiting close to the star can result in tidal heating from the planet itself \citep{Barnes2013}, as the gravity from the star squeezes the planet and creates inner friction. Additionally, many M dwarfs are highly active, with long activity lifetimes \citep{Hawley1996,West2008}.  Magnetic activity from the star can potentially render even a planet well within the habitable zone inhospitable to life.  Flares can irradiate the surface of the planet, or dramatically affect the atmosphere, possibly removing it completely (see for example Segura et al. 2010).

Stellar activity can also inhibit the detection of orbiting planets.  Dark spots on the surface of the star---a result of magnetic activity in the photosphere---can create noise in light curves, and radial velocity ``jitter'': a term which refers to the noise-like structure introduced into the measured radial velocity curve in a systematic or unsystematic way, for example when the wavelength of a stellar line from the integrated stellar surface light is shifted, during the stellar rotation, due to the existence of inhomogeneities on the surface intensity caused by sunspots, plages, or other phenomena moving semi-permanently with the rotating surface or appearing and disappearing during the rotation. In first approximation one could assume that this phenomenon could only look like a planet with an orbital period identical to the stellar rotation period. However, such jitter in the data can cause several different kinds of misinterpretations of the real signal, including false-positive planet identifications, masking the planetary signal, or leading to an erroneous estimate of the planetary parameters. The jitter from stellar activity has the largest probability of dominating the Fourier periodigram when the planetary signal is weak relative to the stellar activity signal, i.e. for small planets and/or stars with large jitter-creating activity, and/or when the obtained data are too sparsely sampled in time.

Spots that persist over timescales significantly longer than the stellar rotation period can mimic a planetary signature, resulting in false-positive ``detections" of planetary companions. \cite{Queloz2001} examined the photometrically variable and magnetically active star HD 166435, which had displayed low-amplitude RV variations with a period of just under 4 days, implying a close-in planetary companion.  However, closer investigation revealed that the RV signal was not coherent beyond durations of approximately 30 days.  \cite{Queloz2001} concluded that the RV variations could be explained by line-profile changes due to dark photospheric spots on HD 166435, and were \textit{not} due to a gravitational interaction with an orbiting planet.  

There are various methods used in attempts to distinguish between spot- and gravitationally-induced RV variations.  Unfortunately, no consistently reliable method has been found.  Bisector analysis is one common method of attempting to classify the origin of observed RV variability.  Spots will distort the absorption line profiles from the star and result in a change in the asymmetry of the lines. These distortions---and the resulting asymmetries---change as the star rotates and the spot crosses from one stellar limb to the other.  The distortions can be quantified with the line bisector span---the difference in bisector value from two different locations in the line profile.  The size of the apparent RV shift resulting from the spots is expected to be correlated with the size of the line bisector span.  Unfortunately, bisector analysis is not always useful in distinguishing the cause of RV variability.  \citet{Prato2008} demonstrated a case where young stars with RV variations that likely resulted from spots did \textit{not} have significant correlations between these RV variations and the bisector spans.  Also, if $v \sin{i}$ is lower that the resolution of the spectrograph, variations in RV and bisector spans will not correlate \citep[see][]{Desort2007}.  

The effects of starspots on RV measurements have been studied by e.g. \cite{Makarov2009}, who derived the amplitude of astrometric, photometric, and RV perturbations caused by a single spot, to compare the sensitivity of astrometric method to the Doppler technique. They concluded that if the ultimate limit of exoplanet detection is defined by intrinsic astrophysical perturbations, the astrometric method is more sensitive than the Doppler technique in this limit.  \cite{Boisse2011} performed simultaneous modeling of stellar activity and planetary parameters in an attempt to remove the RV jitter.  They succeeded in removing up to 90\% of the spot-induced jitter from their test models, but required well-sampled photometry and accurate characterization of the host star in order to distinguish the jitter from the planet-induced RV signal.  \cite{Dumusque2014} developed the software tool \textit{Spot Oscillation And Planet} (SOAP), which simulates the effect of stellar spots and plages on radial velocimetry and photometry.  
\cite{Barnes2011} studied how spot-induced jitter affects planet studies of M dwarfs by using randomly-distributed spot maps to obtain detection thresholds for orbiting planets.  \cite{Reiners2010} investigated spot-induced jitter as a function of observational wavelength as part of a larger study of detection of planets around very low-mass stars using RV measurements, and \cite{Marchwinski2014} used the Sun as a proxy to investigate RV jitter as a function of wavelength by estimating the RV jitter induced by stellar activity on the Sun using solar spectral irradiance measurements.  \cite{Reiners2010} noted that the jitter depends strongly on the properties of the spot, including spot temperature and appearance. 

All of these studies are affected by a limited knowledge of the behavior, amount, and patterns of spots on stellar surfaces, especially those of M dwarfs. Often for M dwarfs, solar models are extrapolated to €œmore active stars€, and random/uniform spot coverage is assumed, which is not necessarily correct.  Doppler Imaging allows ``mapping" of starspots on the stellar surface, but its application to M dwarfs is limited due to the need for fast rotators combined with bright magnitudes, the latter of which limits the number of M dwarfs that can be observed.  New techniques such as Eclipse Mapping through examination of starspot crossing events in photometric data (see \citealt{Sanchis-Ojeda2011}) have revealed more information about spot parameters on M dwarfs stars and make possible exploration of these parameters beyond the limits of Doppler Imaging. 

In this work we combined information from Doppler Imaging, Eclipse Mapping, light curve monitoring, TiO observations, and dynamo models to paint a more realistic picture of M dwarf spots, including spot coverage, spot--to--photosphere temperature contrast ratios, and active regions on the stellar surface, which we discuss in Section 2. Section 3 explains the methods we use to generate line profiles and extract the stellar RV signals.  We first probed the variation in jitter with changes in model parameters such as observational wavelength, stellar temperature, and spot configurations.  We then added a simulated planet to the model, and briefly explored the planet detection limits imposed by stellar jitter.  In Section 4 we investigated the uncertainty contributed by stellar jitter to derived planetary system parameters. 
Finally, our resulting limits on exoplanet detection and characterization due to stellar spot-induced jitter are discussed in Section 5, as well as implications for planet searches, and future work.

\section{M Dwarf Spots}

Unfortunately, despite the ubiquity of small stars, little is still known about M dwarf spots (due, in part, to their low luminosities).  We attempted to model correct spot behavior based on current known constraints and spot information from the literature.  

\begin{figure}
 \includegraphics[width=84mm]{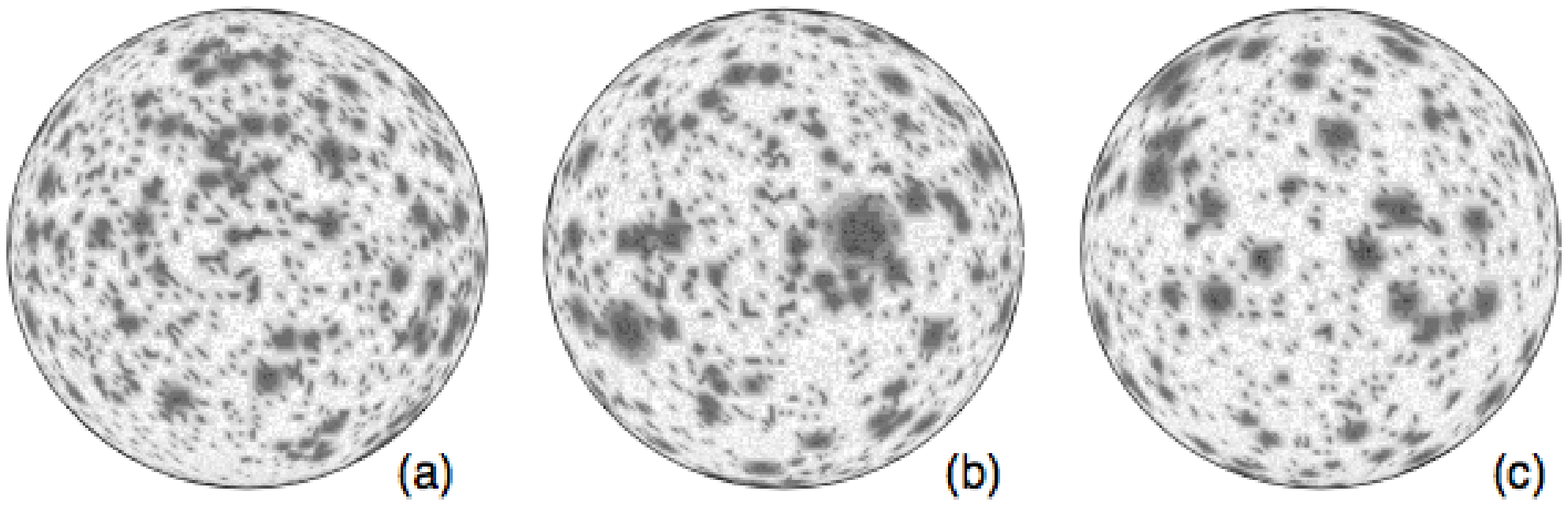} 
  \includegraphics[width=85mm]{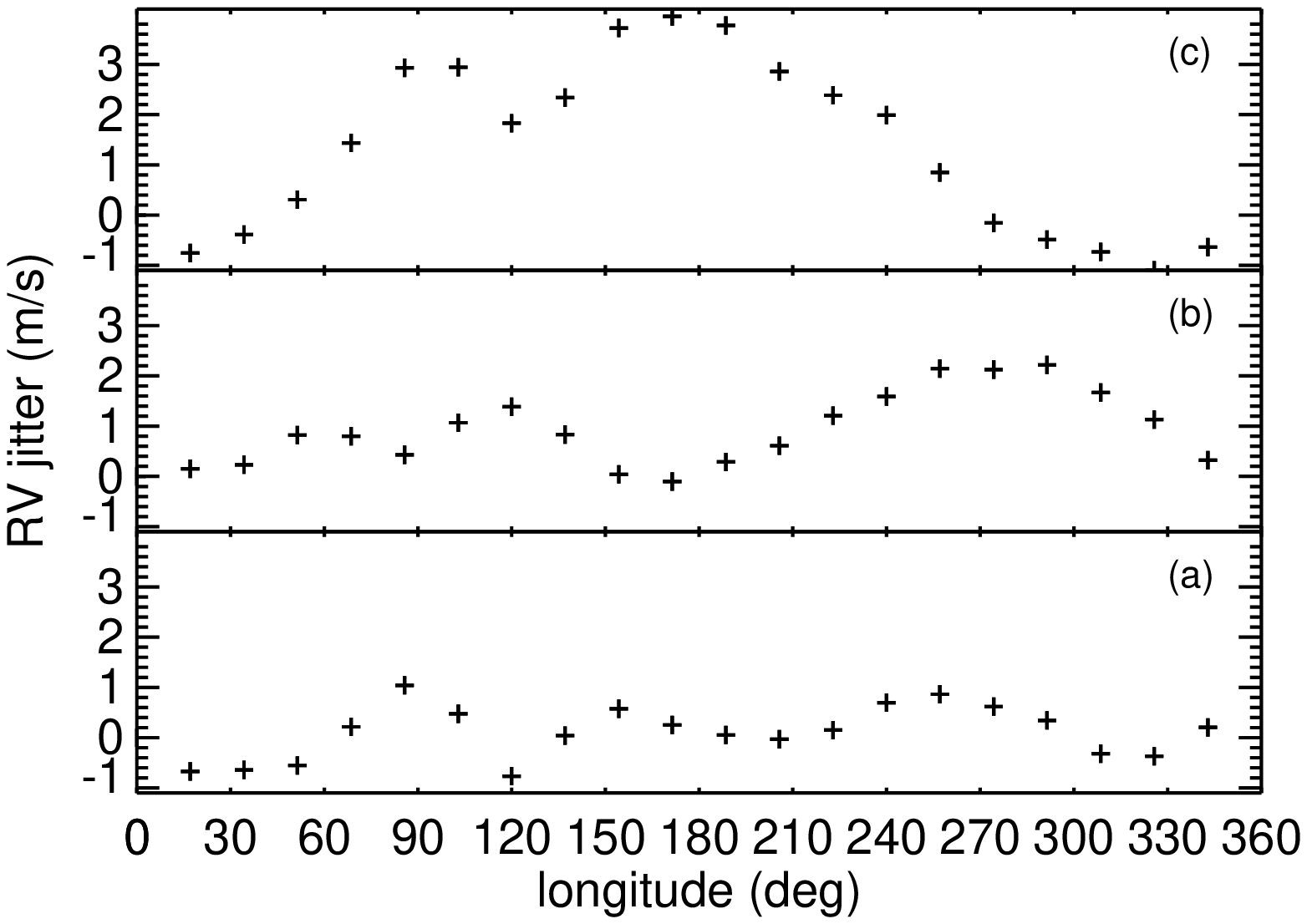} 
\caption{\textit{Top:} Three stellar surfaces with the same photosphere-to-spot temperature contrast ($\Delta T = 500$) and spot filling factor (30\%), but different spot size distributions.  Each was generated by randomly choosing spots from a lognormal distribution of spot radii \citep[following][]{Bogdan1988}, combined with a power law distribution to increase the number of small spots.  The mean value, $\mu$, for the lognormal distribution is the only factor that changes, and corresponds to \textit{(a)}, $\mu_{r} = 0.250$ deg, \textit{(b)}, $\mu_{r} = 0.700$ deg, \textit{(a)}, $\mu_{r} = 1.500$ deg.  \textit{Bottom:} RV jitter generated by the three different spot maps, at 20 observational phases spanning the entire longitude of the stellar surface.}
\label{diff_spot_sizes}
\end{figure}

One manner in which stellar spots are parameterized is by quoting the spot \textit{filling factor}, the percentage of the visible surface area of the star that is covered by dark spots.  Spot filling factors can be measured in various ways, from direct observations in the case of Sunspots \citep[see][]{Solanki1999}, to observations of absorption bands of TiO \citep[e.g. ][]{O'Neal2004}, and to some extent, Doppler Imaging \citep[see ][]{Rice2002}, although this technique will not resolve small spots and thus will underestimate filling factors.  Spot filling factors can also be indirectly inferred by creating model stars with various spot sizes and filling factors and comparing the generated light curves to observational data.  However, simply knowing the spot filling factor is not sufficient, since different configurations of spots will yield significantly different amounts of RV jitter.  An extreme example of this would be to imagine two stars, each with a near 50\% filling factor, where on one star the spots were uniformly distributed across the stellar surface, and on the other star all the spots were grouped on one hemisphere.  These two scenarios would result in vastly different observational signatures.  A less extreme version of this scenario is illustrated in Figure \ref{diff_spot_sizes} where three simulated stellar surfaces are shown.  Each has the same spot filling factor, but a different size distribution of spots, resulting in a differing level of homogeneity of the spot coverage for each star.  The derived RV jitter from each surface shows how even a small difference in spot size can result in a significant (in this case, a factor of almost 3) change in spot-induced jitter.   

To correctly model M dwarf spots, we made use of constraints on spot size, temperature, location, and behavior from various observational methods such as Doppler Imaging, Zeeman Doppler Imaging, investigation of light curve modulation, and eclipse mapping, as well as modeling of M dwarf dynamo behavior.  Evidence for both random, uniform spottedness as well as spots concentrated in active latitudes and/or longitudes has been observed.

\subsection{Spot size and distribution}

Doppler Imaging (DI) operates on the same principle as spot--induced jitter (dark spots on the stellar surface result in an anomaly in the observed spectrum), by ``inverting" the spectrum to extract the surface distribution of spots that created it. This is one of the few ways to obtain a ``map" of a stellar surface, although the solution is degenerate, so results should be taken with caution.  Due to their low-luminosities, fewer than ten K and M dwarfs have been successfully imaged using the DI technique.  Of these, the results are mixed: 

Some low--mass stars have shown evidence of high--latitude (or polar), long--lived spots or spot groups, that appear at some epochs and are not present at others.  For example, AB Dor, a rapidly--rotating, ZAMS K dwarf displayed this type of spot behavior in 1993--1994 \citep{Jeffers2007}, but not in 1989 \citep{Kuerster1994}, and BO Mic displayed similar high--latitude spots in 2002 \citep{Barnes2005a} but not in 1998 \citep{Barnes2001}.  This behavior could be indicative of stellar cycles on these stars.  

The rapidly--rotating M dwarfs HK Aqr (M1, P=0.4 days) and EY Dra (M2, P=0.5 days) show spots at low latitudes or with no strong latitude dependence \citep{Barnes2001}, though based on photometric observations EY Dra has shown indication of active longitudes \citep{Vida2010} where spots preferentially appear.

Further evidence for non-uniform spot distribution is provided by the Eclipse Mapping technique, where properties of starspots can also be inferred through observation of anomalies in high signal--to--noise, high--cadence light curves of  planetary transits \citep[see][]{Pont2007, Wolter2009, Sanchis-Ojeda2011}. This has some advantage over Doppler Imaging since the stars do not need to be rapidly rotating and it is possible to obtain information about less luminous M dwarfs than would be possible through Doppler imaging.  In eclipse mapping, the eclipsing body ``scans" the face of its companion and passes over dark spots, and the resulting changes in light level can be used to reconstruct the surface brightness distribution of the eclipsed face. \cite{Wolter2009}, \cite{Silva-Valio2010} and \cite{Silva-Valio2011} found numerous Spot Crossing Events (SCEs) in the light curve of CoRoT-2, a rapidly--rotating (P=4.5 d) G7V star.  They found that an average of five SCEs occurred per transit, and that the transited region had a spot filling factor of 10 -- 20 percent.  

Eclipse mapping is still difficult for M dwarfs because it requires a transiting object to scan the face of the star.  In the Kepler catalog, there are approximately 4000 M dwarfs \citep{Dressing2013}. Of these, so far only one has been identified that has both a transiting planet and a corresponding anomaly in the light curve indicative of an SCE: KOI 1152.  The dearth of M dwarf SCEs could imply either low contrast ratios between spot temperatures and photospheric temperatures, or stars that have high filling factors of more or less homogeneously distributed spots (or simply that the probability of a planet crossing a spotted region is low).  KOI 1152 indicates a large spot size of approximately 20 degrees in diameter, and a low temperature contrast ratio of about 100 degrees difference between the spot temperature and the photospheric temperature of the star (Sanchis-Ojeda, private communication).  \cite{Sanchis-Ojeda2011} also identified spot-crossing anomalies on the K4V star HAT-P-11 which preferentially occurred at two specific phases of the transit. They interpreted these as intersection points between the transit chord and active latitudes of the host star.  Based on these data, it would make sense to model both homogeneously distributed, high filling factor spot configurations as well as large spots and/or spot concentrations in specific active regions.  

Analysis of the broad-band light curves of magnetically active stars seems to yield indirect evidence of a more random, uniform distribution of spots.  The amount of rotational modulation in the stellar light curve depends on the degree of axisymmetry of the starspot distribution. Thus, observation of a \textit{lack} of rotational modulation in the light curve could be indicative of either a spotless surface or a very uniform spot coverage that does not significantly alter the luminosity of the visible surface with rotation.  \cite{Jackson2012} examined magnetically active low-mass members of NGC 2516.  While some displayed small amplitude modulations with rotation, they noted that almost half of the monitored stars had no detectable light-curve modulation at all, despite showing similar chromospheric emission levels (indicative of magnetic activity), having a similar distribution of equatorial rotation velocities, and showing no significant difference in distribution of positions on a color-magnitude diagram.  They suggested it should be possible to have high spot filling factors and still display small light-curve amplitudes if the stars were covered by many small, randomly placed spots with angular diameters of $\sim2\degree$.

\cite{Jackson2013} further examined this proposal through simulations of  low-mass magnetically active stars with  randomly orientated stellar spin axes and cool starspots of a characteristic scalelength randomly distributed across the stellar surfaces.  They compared the models with their observations of NGC 2516 M dwarfs and found that the best-fitting starspot angular scalelength is about 3.5 degrees, although this is dependent on the the assumed spot temperature ratio and filling factor (i.e. the model cannot be constrained by the light curve data alone).

\cite{Savanov2012}, however, found evidence of two active longitudes on the fully-convective M dwarf LHS 6351 using light curve monitoring, similar to the \citet{Vida2010} active longitudes. 

Zeeman Doppler Imaging (ZDI) seems to further support the idea of a more uniform spot distribution.  ZDI allows mapping of the magnetic topography of the star by detecting spectral line splitting due to the Zeeman effect.  Like regular Doppler Imaging, the result is degenerate, and the reconstruction of the magnetic geometry yields only the simplest solution that fits the observed data.  There is evidence from ZDI  of more axisymmetric magnetic field topology in fully-convective stars \citep[approximately M4 and later, see][]{Morin2008, Morin2010}.

Dynamo simulations of M dwarf stars have been carried out, though the resulting consequences on spot formation (and location) are often not mentioned.  \cite{Browning2008} ran three-dimensional nonlinear magnetohydrodynamic simulations of the interiors of fully convective M dwarfs to ascertain if a magnetic dynamo could be produced without the shear caused by the interface between convective and radiative zones as in larger stars.  Although \cite{Browning2008} concluded that magnetic fields can be generated in fully-convective stars, and that the fields would have a strong axisymmetric component, they stopped short of speculating on the resulting spot formation on the stellar surface.   

\cite{Granzer2000} performed simulations of the dynamics of magnetic flux tubes in low-mass ZAMS stars (with masses between 0.4\,M$_{\odot}$ and 1.7\,M$_{\odot}$), and calculated the spot emergence latitudes at different rotation rates. They found that  for a 0.4\,M$_{\odot}$ star, spots form at least 20 deg from the equator even at the slowest rotation rates.  At higher rotation speeds, they found spots concentrated between $\sim$50 to 70 degrees in latitude.

\subsection{Spot temperatures}
Spot radius and temperature are highly correlated when fitting light-curve (LC) and eclipse-mapping (EM) data.  Thus, a knowledge of spot temperatures would be ideal, although difficult to obtain.  Spot temperatures can also be obtained using molecular band head modeling (MB), line depth ratios (LDR), and DI.  DI seems to indicate that cooler stars show a lower temperature difference between photospheric temperature and spot temperature \cite[][]{Strassmeier2011}. \cite{Berdyugina2005} showed an average temperature difference for M dwarfs of just 200\,K, while for larger stars $\Delta$T values were observed up to 2000\,K. 

\cite{Gray1991} demonstrated the LDR technique of using two nearby spectral lines of different temperature sensitivity to obtain spot temperatures. This method was employed by e.g. \cite{Catalano2002}
and \cite{Frasca2005} who obtained a range of 400--1000 K for $\Delta$T for a number of giant stars.  

Due to their low temperatures, cool dwarfs exhibit many TiO bands which can be used to derive spot temperatures, using the MB technique.  \cite{O'Neal2004} obtained relative spot temperatures for five G and K dwarfs using the TiO band heads at 705 nm and 886 nm and found $\Delta$T values in the range of 1000--2000 K.  

\cite{Rice2010} obtained two separate sets of Doppler images of the WTTS V410 Tauri star, from atomic lines and from the TiO 705.5-nm lines respectively, and found that the TiO-based temperatures appeared to be cooler, with an average temperature difference of 105\,K between the two techniques. This could partly be a result of the difficulty in determining the correct continuum in the forest of TiO lines, however they concluded that atomic-line DI is prone to underestimate spot temperatures on a level of about 100\,K for a 4500\,K photospheric temperature.

\begin{figure}
 \includegraphics[width=84mm]{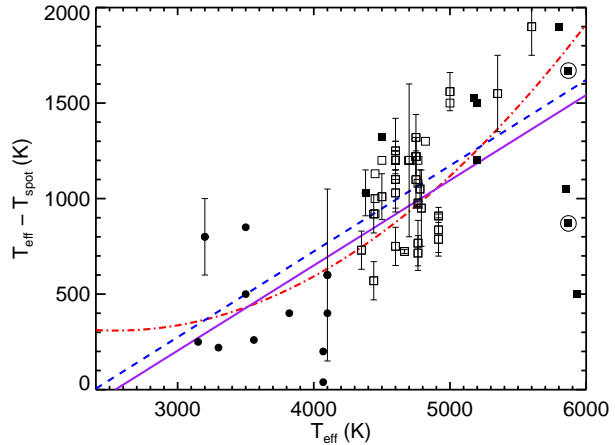} 
\caption{Spot temperature contrasts from published values (see Tables \ref{spot_temps_giants} and \ref{spot_temps_dwarfs}).  Filled points indicate dwarfs and open points are giants. M dwarfs are identified by filled circles.  Contrast values for the Sun are indicated by circled points and represent the umbral (higher) and penumbral (lower) temperature contrasts.  Only the umbral contrast in included in the fits.  Blue dashed line is linear fit to all points.  Solid purple line is linear fit to dwarf stars only.  Red dot-dash line is polynomial fit to dwarf stars only. Linear fit lines have limited use at predicting temperature contrasts at low photospheric temperatures due to the fact that they reach zero in the T = 2400 - 2600\,K range.}
\label{fig:contrast_plot}
\end{figure}

Figure \ref{fig:contrast_plot} shows spot temperature contrasts from the literature, including both giant stars (Table \ref{spot_temps_giants}) and dwarfs (Table \ref{spot_temps_dwarfs}).  Only twelve data points (for eight individual stars) exist for M dwarfs.  We attempted to fit a line to these points but the resulting line suffered from small numbers, and had a slope opposite to the linear and polynomial fits that include more points, likely indicating that more data are needed before such a fit will realistically reflect the change in spot contrast with photospheric temperature of M dwarf stars.  Linear fit lines are also of limited use for predicting temperature contrasts at the low photospheric temperatures of M dwarfs due to the fact that the lines reach zero in the T = 2400 - 2600\,K range. Since the data are so sparse and the error bars so large, it seems the most we can currently conclude about M dwarf spot temperature is that $\Delta T$ seems to be in the range of 200 - 600\,K, with a couple stars showing higher contrasts of up to $\sim$800\,K.

\begin{table*}
\begin{minipage}{126mm}
 \caption{Giant star spot temperature contrasts from literature. Where available, photometric variation ($\Delta$V) and spot filling factor (ff) are also included. The method used to obtain the spot temperatures is indicated as: molecular band head modeling (MB), light curve monitoring (LC), line depth ratios (LDR), doppler imaging (DI), or eclipse mapping (EM). }
 \label{spot_temps_giants}
 \begin{tabular}{@{}lccccccccc}
  \hline
  Star 
  	& SpT
    	& $T_{phot}$ 
  	& $T_{spot}$
        & $\Delta T $
        & $\Delta V $
        &ff (\%)
        & method
        & reference
          \\
  \hline    

HD199178    &G5\,III    &5350    &3800    &1550$\pm$200    &\dots  &29    &MB
    &1*\\
EI Eri    &G5\,IV    &5600    &3700    &1900$\pm$150    &\dots    &36    &MB
    &1*\\
$\lambda$ And    &G8\,III    &4750    &3650    &1100$\pm$150    &\dots    &23   &MB
    &1*\\
     &G8\,IV    &4780    &3730    &1050$\pm$100    &\dots    &17   &LC &4\\
HK Lac    &K0\,III    &4790    &3840    &950$\pm$200    &0.14    &\dots    &LC
    &2*\\
     &K0\,III    &4765    &3788    &977$\pm$10    &0.37    &34  &LDR    
     &3*\\
     &K0\,III    &4600    &3520    &1080$\pm$100    &\dots    &8    &LC
    &4\\
     &K0\,III    &4765    &3565    &1200$\pm$100    &0.74    &\dots    &LC &5\\
    &K0\,III    &4765    &3998    &767$\pm$119    &0.07    &13  &LDR &15\\
    &K0\,III    &4765    &4050    &715$\pm$91    &0.07    &14  &LDR &15\\
    
XX Tri    &K0\,III    &4820    &3500    &1300    &0.63    &20    &DI
    &6*\\
     &K0\,III    &4750    &3420    &1320$\pm$120    &\dots    &35    &MB
    &7*\\
    
HU Vir    &K0\,IV    &5000    &3440    &1560$\pm$100    &\dots    &44    &MB
    &1*\\
      
UX Ari    &K0\,IV    &5000    &3500    &1500    &\dots    &48    &MB    &8*\\
    &K0\,IV    &4780    &3360    &1420$\pm$100    &\dots    &8    &LC    &4\\
        
  LX Per   &K0\,IV    &4780    &4050    &730$\pm$100    &\dots    &5    &LC &4\\
    
    AR Lac    &K0\,IV    &4700    &3500    &1200    &0.04    &\dots    &LC    &9*\\
$\sigma$ Gem    &K1\,III    &4600    &3850    &750$\pm$100    &\dots    &33    &MB
    &1*\\
    &K1\,III    &4440    &3870    &570$\pm$100    &\dots    &8    &LC &4\\
    
DM UMa    &K1\,III    &4600    &3570    &1030$\pm$100    &\dots    &35    &MB
    &1*\\
     &K1\,III    &4500    &3450    &1010$\pm$120    &\dots    &42    &MB
    &7*\\
    
SZ Psc    &K1\,IV    &4700    &3500    &1200$\pm$400    &\dots    &\dots    &LC    &10*\\

HR1099    &K1\,IV    &4700    &3500    &1200    &\dots    &40    &MB    &8*\\
     &K1\,IV    &4700    &3500    &1200    &0.09    &\dots    &LC
    &9*\\
    
HR7275   &K1\,IV    &4600    &3400    &1200$\pm$100    &\dots    &9    &LC  &4\\
 &K2\,III    &4600    &3500    &1100$\pm$150    &\dots    &27    &MB
    &1*\\ 
    
    IM Peg     &K2\,II    &4450    &3450    &1000    &\dots    &20    &DI    &12*\\
    &KI\,V    &4400    &3270    &1130$\pm$35    &0.38    &15    &LC  &13\\
      &K2\,II    &4666   &4218    &448$\pm$126    &0.17    &11    &LDR   &15\\
       &K2\,II    &4666    &4269    &397$\pm$140    &0.17    &12    &LDR    &15\\
       &K2\,III    &4666    &3943    &723$\pm$10    &0.20    &32
    &LDR    &3*\\
       &K2\,III    &4440    &3520    &920$\pm$100    &\dots    &7    &LC    &4\\
       
II Peg    &K2\,IV    &4600    &3400    &1200$\pm$100    &0.42    &37
    &LC    &2*\\
     &K2\,IV    &4750    &3530    &1220$\pm$100    &0.16    &50
    &MB    &1*\\
     &K2\,IV    &4750    &3530    &1220$\pm$100    &0.23    &43
    &MB    &1*\\
     &K2\,IV    &4500    &3300    &1200    &0.25    &\dots    &LC    &9*\\
     &K2\,IV    &4600    &3500    &1100    &0.25    &20    &DI
    &11*\\
    &K2\,IV    &4350    &3620    &730$\pm$100    &\dots   &\dots    &LC &4\\

IN Vir    &K2\,IV    &4600    &3350    &1250$\pm$170    &\dots    &40    &MB
    &7*\\
VY Ari    &K3\,IV    &4916    &4007    &909$\pm$10    &0.41    &41
    &LDR    &3*\\
     &K3\,IV    &4600    &3400    &1200$\pm$50    &0.28    &15
    &LC    &14*\\
     &K3\,IV    &4600    &3400    &1200$\pm$50    &0.20    &12
    &LC    &14*\\
     &K3\,IV    &4600    &3400    &1200$\pm$50    &0.10    &15
    &LC    &14*\\
     &K3\,IV    &4916    &4080    &836$\pm$118    &0.15    &15   &LDR  &15\\
     &K3\,IV    &4916   &4129    &787$\pm$88    &0.15    & 16   &LDR  &15\\
  \hline
  
  \end{tabular}

  \medskip
  References: 1, \cite{O'Neal1998};
  2, \cite{Vogt1981};
  3, \cite{Catalano2002};
  4, \cite{Poe1985};
  5, \cite{Olah1997};
  6, \cite{Strassmeier1999};
  7, \cite{O'Neal2004};
  8, \cite{O'Neal2001},
  9, \cite{Rodono1986};
  10, \cite{Eaton1979};
  11, \cite{Berdyugina1999a};
  12, \cite{Berdyugina2000};
  13, \cite{Padmakar1999};
  14, \cite{Strassmeier1992};
 15, \cite{Frasca2005};
   \begin{tablenotes}
            \item [1]  * indicates that these values are also included in the \cite{Berdyugina2005} review. 
                     \end{tablenotes}

    \end{minipage}
 \end{table*}

\begin{table*}
\begin{minipage}{126mm}
 \caption{Dwarf star spot temperature contrasts from literature. Where available, photometric variation ($\Delta$V) and spot filling factor (ff) are also included.}
 \label{spot_temps_dwarfs}
 \begin{tabular}{@{}lccccccccc}
  \hline
  Star 
  	& SpT
    	& $T_{phot}$ 
  	& $T_{spot}$
        & $\Delta T $
        & $\Delta V $
        &ff (\%)
        & method
        & reference
          \\
  \hline    
  
Sun    &G2\,V    &5870    &4200    &1670    &    & 0.03 - 0.30    &    &25\\
   &G2\,V    &5870    &5000    &870    &    &    &    &25\\
  \hline
  
EK Dra    &G2\,V    &5930    &5400    &500    &\dots    &6    &LC    &16*\\
     &G2\,V    &5850    &4800    &1050    &0.08    &11    &DI &17*\\
     &G2\,V    &5830    &3800    &2030    &\dots    &40    &MB    &7*\\
HD307938    &G2\,V    &5800    &3900    &1900    &0.06    &13    &DI+LC &18*\\
AB Dor    &K0\,V    &5200    &4000    &1200    &0.05    &12    &LC  &19*\\
     &K0\,V    &5200    &3700    &1500    &0.12    &5    &LC &19*\\
LQ Hya    &K2\,V    &5175    &3650    &1525    &\dots    &45    &MB    &8*\\
V833 Tau    &K4\,V    &4500    &3175    &1325    &\dots    &45    &MB    &8*\\
EQ Vir    &K5e\,V    &4380    &3350    &1030$\pm$120    &\dots    &45    &MB &7*\\
BY Dra    &M0\,V    &4100    &3500    &600$\pm$450    &0.08    &34 &LC    &2*\\
     &M0\,V    &4100    &3700    &400    &\dots    &60    &LC    &20*\\
     &M0\,V    &4100    &3500    &600    &0.10    &\dots    &LC    &9*\\
YY Gem    &M0\,V    &3820    &3400    &400    &\dots    &\dots    &LC    &21*\\
KOI 1152    &M1\,V    &4069    &3869    &200    &\dots    &\dots    &EM    &26\\
  &M1\,V    &4069    &4030    &39    &\dots    &\dots    &EM    &29\\
HHJ 409    &M\,V    &3560    &3300    &260    &0.08    & 13   &LC    &27\\
AU Mic    &M2e\,V    &3500    &2650    &850    &0.10    &\dots    &LC    &9*\\
     &M2e\,V    &3500    &3000    &500    &0.32    &10    &LC &22*\\
EV Lac    &M4e\,V    &3300    &3080    &220    &0.10    &7    &LC  &23*\\
BPL129    &M4\,V    &3200    &2400    &800$\pm$200    &\dots    &4    &LC   &24\\    
LHS 6351    &M\,V    &3150    &2900    &250    &\dots    &1   &LC   &28\\    
    \hline
  
  \end{tabular}

  \medskip
  References: 
  2, \cite{Vogt1981};
  7, \cite{O'Neal2004};
  8, \cite{O'Neal2001},
  9, \cite{Rodono1986};
  16, \cite{Dorren1994};
  17, \cite{Strassmeier1998};
  18, \cite{Marsden2005};
  19, \cite{Amado2001};
  20, \cite{Chugainov1976};
  21, \cite{Torres2002};
  22, \cite{Torres1973};
  23, \cite{Abranin1998};
  24, \cite{Scholz2005};
  25, \cite{Berdyugina2005};
  26, Sanchis Ojeda, private communication;
  27, \cite{Terndrup1999};
  28, \cite{Savanov2012};
  29, \cite{Varga2014} 
 
 \begin{tablenotes}
            \item [1]  * indicates that these values are also included in the \cite{Berdyugina2005} review.
                      \end{tablenotes}

    \end{minipage}
 \end{table*}

\section{Simulating Radial Velocity Data}
We use these observations and simulations of spot behavior as constraints on our modeled M dwarf spot patterns.  We create cases with random, ``uniform" spot coverage (i.e. no preferential latitude or longitude spot emergence) as well as cases with active latitude regions following the \cite{Granzer2000} and  \cite{Vida2009} models.  

 We developed the SPOTSS (Simulated Patterns Of Temperatures on Stellar Surfaces) code to simulate stellar spot patterns  \citep[see][hereafter Paper I, for more details]{Korhonen2014}. Using SPOTSS we can generate either random spots, or spots constrained to a certain active latitude and/or longitude region.  The spots created by SPOTSS have a umbral region which has a temperature defined as the spot temperature, and a penumbral region, with a temperature of halfway between the spot temperature and photospheric temperature.  Spots are generated with a lognormal size distribution \citep[the size distribution observed for Sunspots, see e.g.][]{Bogdan1988}, combined with a power law to slightly increase the number of small spots for M dwarfs due to observations by \cite{Jackson2012} which indicate M dwarfs have numerous small spots covering the surface.
 
 In all our simulations we used a stellar rotation period of P = 7.7 days.  Paper I examines the effects of $v\sin{i}$ on the measured jitter.  We are interested here in exploring the effect of spot configuration on jitter and thus we keep a constant $v\sin{i}$ for all our ``observations." A 7.7-day period is a reasonable choice for M dwarfs, and was chosen because it is in between the very active and the non-active cases.  Since the choice of activity level (filling factor) and spot configuration have a much higher effect on the jitter than the $v\sin{i}$, using P=7.7\,d for all our measurements does not significantly bias our results.

\subsection{RV Jitter from spots}
Radial velocity measurements were generated from our simulated stellar surfaces using DIRECT7 \citep{Piskunov1990, Hackman2001} and DEEMA (Detection of Exoplanets under the Effect of Magnetic Activity) codes. For M dwarf photospheric temperatures we used the MARCS models \citep{Gustafsson2008}, with solar metallicity, and $log{g} = 5$, a temperature grid of 2500 - 3900\,K, with a 200\,K step size, and 17 limb angles.  DIRECT7 calculates synthetic spectra by integrating spectral line profiles over the entire visible stellar surface at each observational phase, using an evenly spaced wavelength grid. The $v \sin{i}$ and inclination values are taken into account to calculate which part of the stellar surface is visible and the rotational broadening of the spectral lines.  The number of observational phases is evenly distributed over the length of the observing run.  DEEMA obtains radial velocity (jitter) measurements by cross-correlating the calculated line profiles with the line profile from the spectrum generated at the first observational phase.  See Paper I for a more detailed description of the codes and process.

\begin{figure*}
\begin{minipage}{168mm}
 \includegraphics[width=84mm]{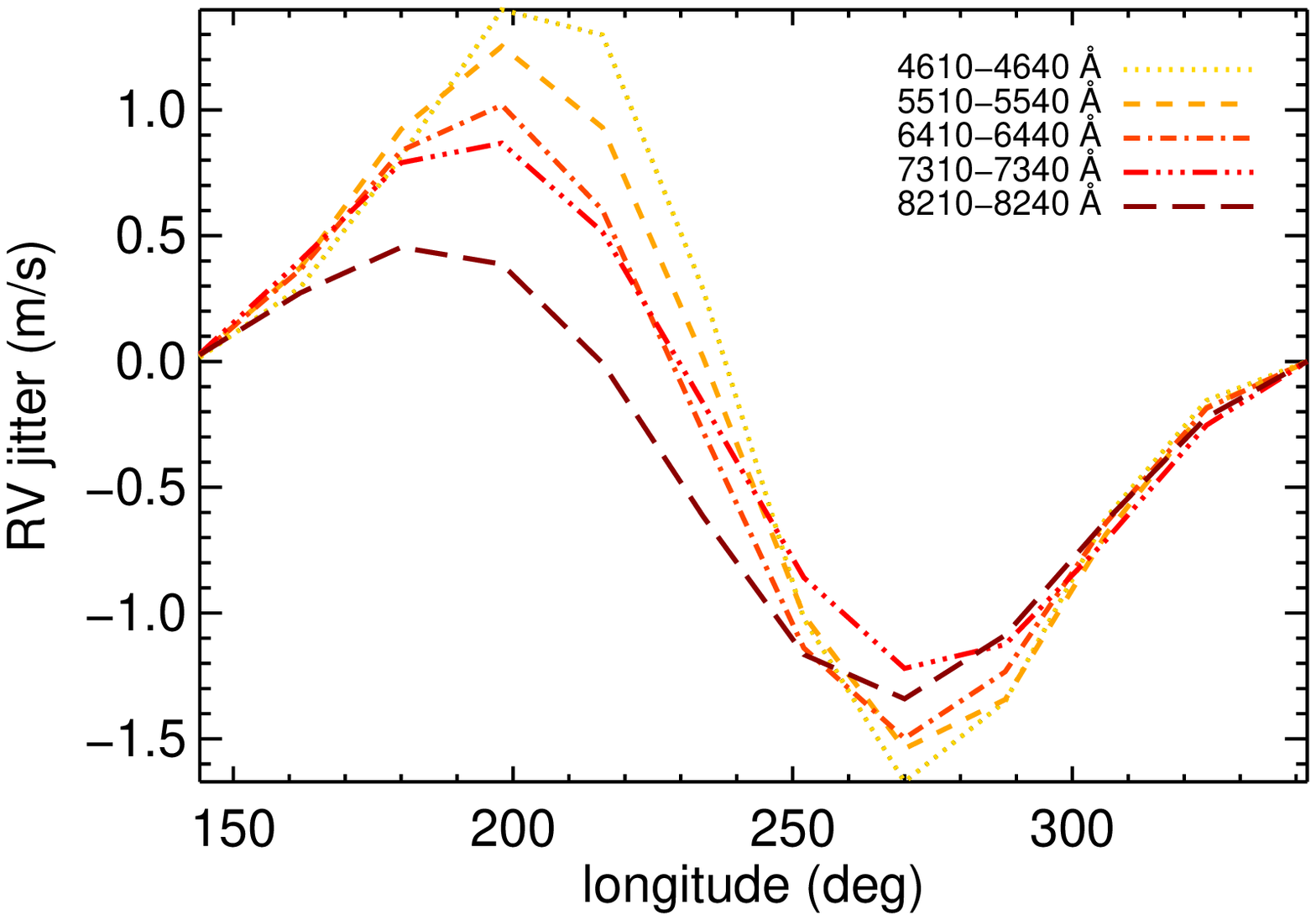} 
  \includegraphics[width=84mm]{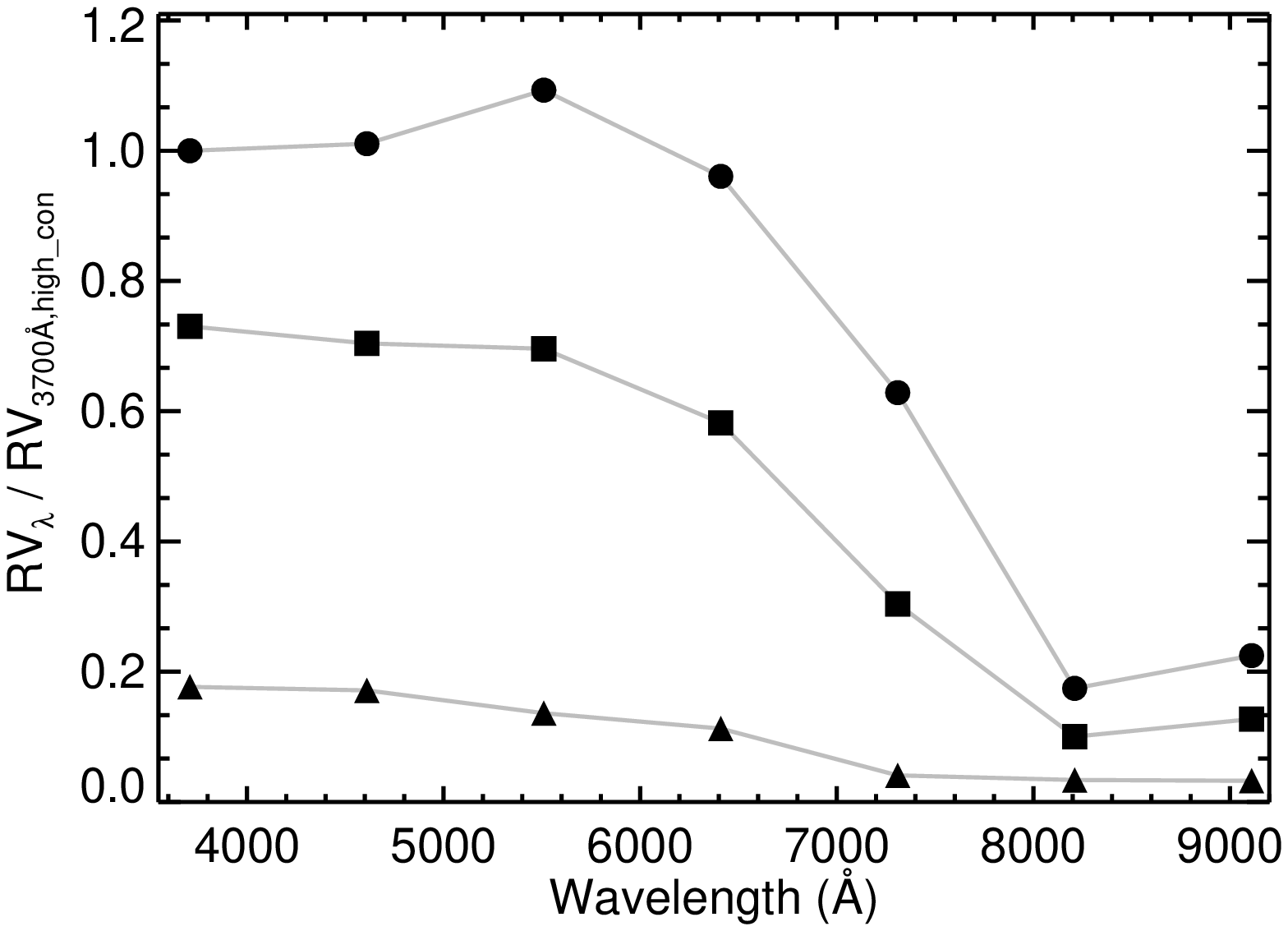} 
\caption{\textit{Left:} RV jitter generated by a single, 3-degree spot at five different observational wavelength ranges: 4610-4640\,\AA, 5510-5540\,\AA, 6410-6440\,\AA, 7310-7340\,\AA, and 8210-8240\,\AA. \textit{Right:} Semi-amplitude of RV jitter (normalized to the jitter measured at 3700\,\AA) as a function of wavelength, at thee different photosphere-to-spot temperature contrasts:  $\Delta T = 100$ (triangles), $\Delta T = 500$ (squares), and $\Delta T = 900$ (circles).  RV jitter is higher at all wavelengths for higher $\Delta T$. Redder wavelengths reduce RV noise from spots, since the contrast between the photosphere and the spot is lower at these wavelengths, although this effect is less dramatic at low contrast ratios.}
\label{RV_wl_curves}
\end{minipage}
\end{figure*}

Figure \ref{RV_wl_curves} shows the RV jitter from calculations using seven 30\,\AA\ wavelength intervals (i.e. 4610-4640\,\AA, 5510-5540\,\AA, etc) ranging from 3710\,\AA\ to 9110\,\AA. Three different photosphere-to-spot temperature contrasts were investigated: a low-contrast case of $\Delta T$ = 100K, a medium contrast of $\Delta T$ = 500K, and a high contrast of $\Delta T$ = 900K.  Our results confirmed previous studies, which showed that jitter decreases with increasing wavelength; redder wavelengths have a lower spot-induced jitter.  At high contrast, we saw a slight increase going from 8210\,\AA\ to 9110\,\AA, similar to \cite{Barnes2011}. Although it is tempting to conclude from our analysis that the best wavelength for planet hunting around M dwarfs would be in the IR, it is also important to chose a bandpass with a large amount of spectral information for the cross-correlation to yield the most precise results.  

\begin{figure*}
\begin{minipage}{168mm}
 \includegraphics[width=84mm]{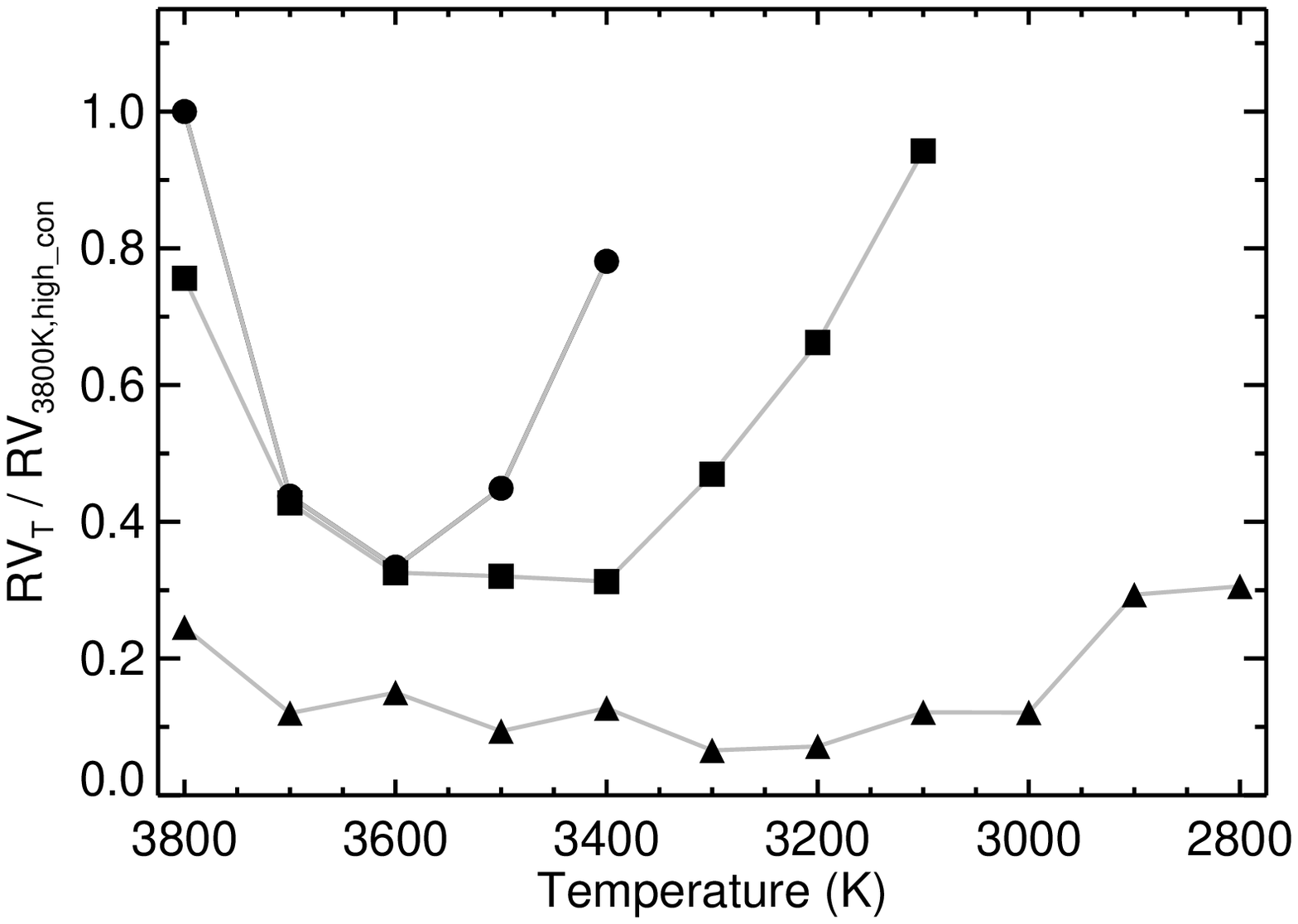} 
  \includegraphics[width=84mm]{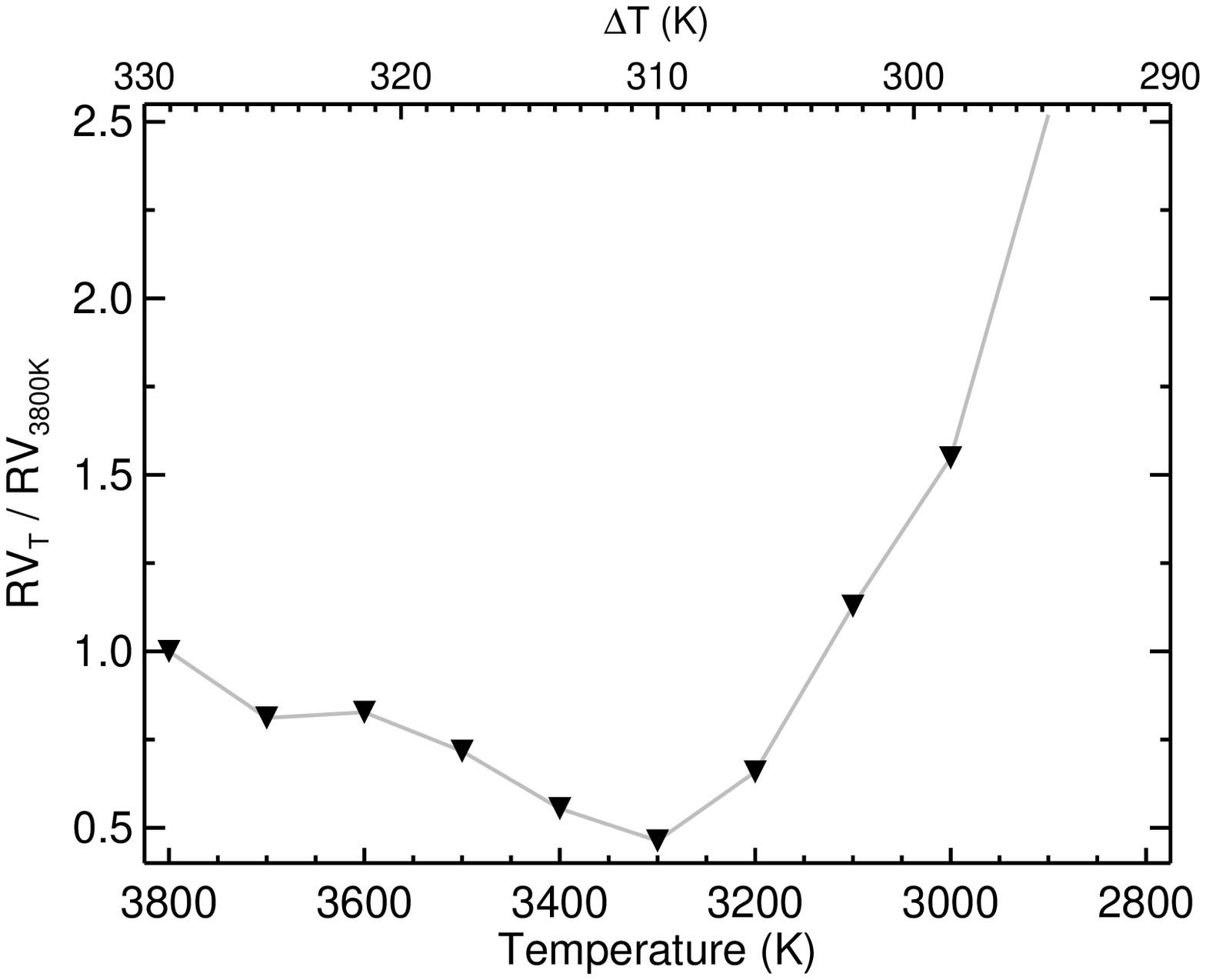} 
\caption{\textit{Left:} Semi-amplitude of RV jitter generated by a single, 10-degree spot at a range of photospheric temperatures from $T_\mathrm{eff}$ = 3800\,K to 2800\,K, corresponding to approximate spectral types M0 -- M8.  Three curves representing low ($\Delta T = 100$, triangles), medium ($\Delta T = 500$, squares), and high ($\Delta T = 900$, circles) contrasts are shown. Jitter is normalized to the jitter measured at $T_\mathrm{eff}$=3800\,K, $\Delta T = 900$.  \textit{Right:} Semi-amplitude of RV jitter as a function of stellar temperature for a constant photosphere-to-spot temperature ratio given by $T_\mathrm{spot}/T_\mathrm{eff}$ = 0.9 (e.g., for $T_\mathrm{eff}$ = 3800\,K, $T_\mathrm{spot}$ = 3420\,K). Jitter is normalized to the jitter measured at $T_\mathrm{eff}$=3800\,K.}
\label{RV_SpT}
\end{minipage}
\end{figure*}

We also tested the RV jitter as a function of stellar photospheric temperature (spectral type).  We investigated the same low, medium and high $\Delta T$ values, as before, but we also added a 4th case where we used $T_\mathrm{spot}$ = 0.9\,$T_\mathrm{eff}$ for the spot temperature (e.g., for $T_\mathrm{eff}$ = 3800\,K, $T_\mathrm{spot}$ = 3420\,K) in order to keep the contrast ratio constant as we changed spectral type (see Figure \ref{RV_SpT}), since the photospheric temperature was changing and thus a constant $\Delta T$ would not result in a constant $T_\mathrm{eff}/T_{spot}$ ratio.  We found that the RV jitter would first decrease, then reach a minimum, and then increase again with decreasing photospheric temperature (i.e. increasing spectral type).  The approximate spectral type of this minimum was dependent on contrast ratio.  At low contrast, the effect was less extreme. It is unclear whether this effect is real or could be caused by possible problems with creating synthetic spectra at lower temperatures.

\begin{figure}
 \includegraphics[width=84mm]{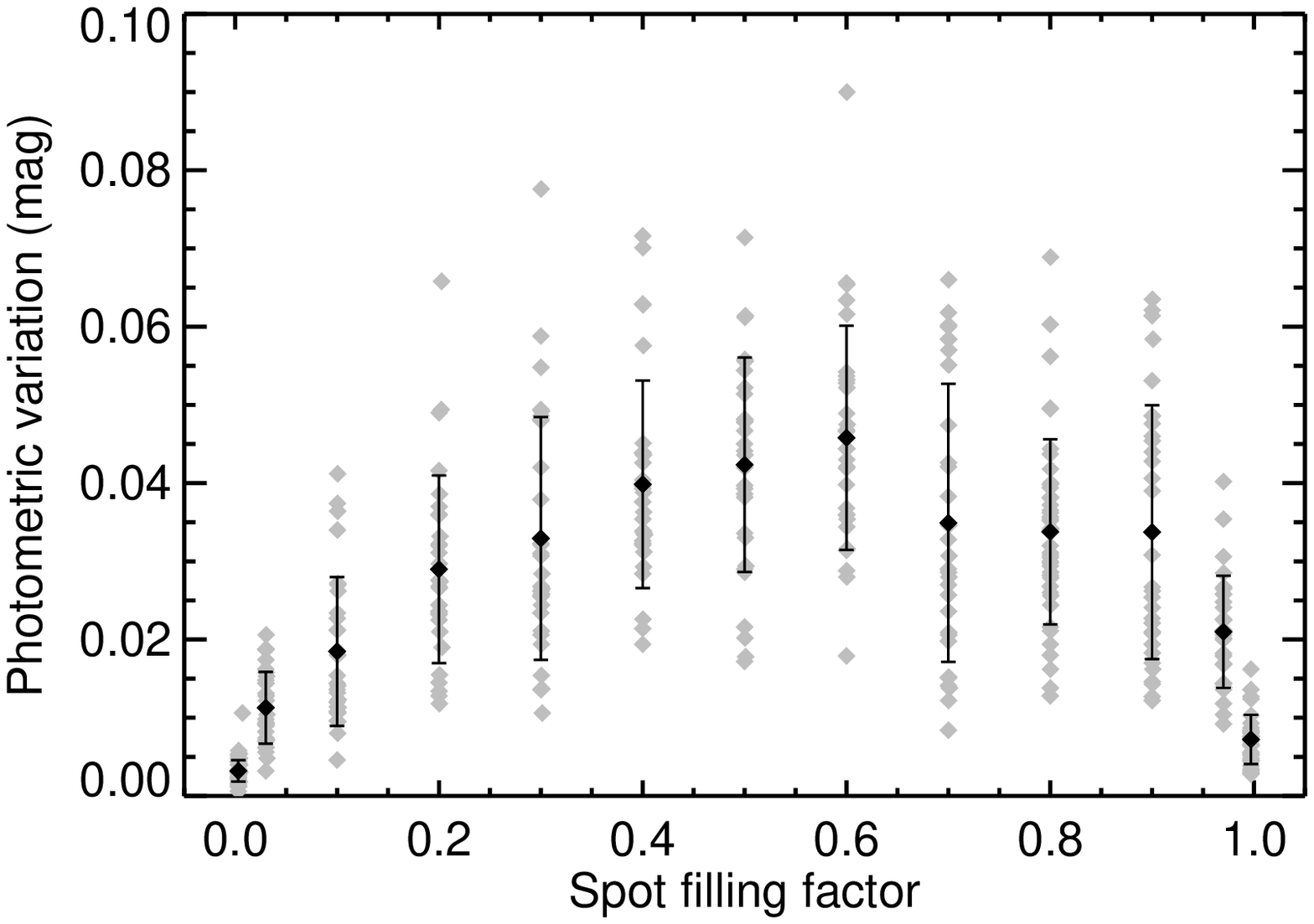} 
\caption{Observed V photometric variation of star with spot filling factor. Grey diamonds represent each individual measurement of jitter from a unique stellar surface consisting of randomly distributed spots with a given filling factor.  Black diamonds correspond to the mean value of the measured jitter at each filling factor, and error bars represent standard deviation calculated from all the measurements at each filling factor.  Jitter increases to a maximum which is reached at approximately 50\% spot coverage, and then decreases again as the spots dominate the stellar surface.}
\label{Vmag_ff}
\end{figure}

With the exception of the Sun, we cannot directly observe stellar spot filling factors. Therefore, we used DIRECT7 to investigate the photometric behavior calculated from the input spot configuration maps and how the variation in magnitude relates to the filling factor, and the resulting RV jitter.  Figure \ref{Vmag_ff} shows the variation in $\Delta V_{mag}$ as a function of spot filling factor, from 0.3\% to 99.7\% filling factor.  The photometric variation increases with increasing spot filling factor to a peak around a filling factor of 50\%, and then decreases again.  This is expected since high filling factors of dark spots would simply appear as bright spots on a dark surface.  The slope of this curve, however, is dependent on the size distribution of the spots, since a few large spots would make a larger difference in the photometric variation than many small, evenly-spaced spots.

\begin{figure}
 \includegraphics[width=81mm]{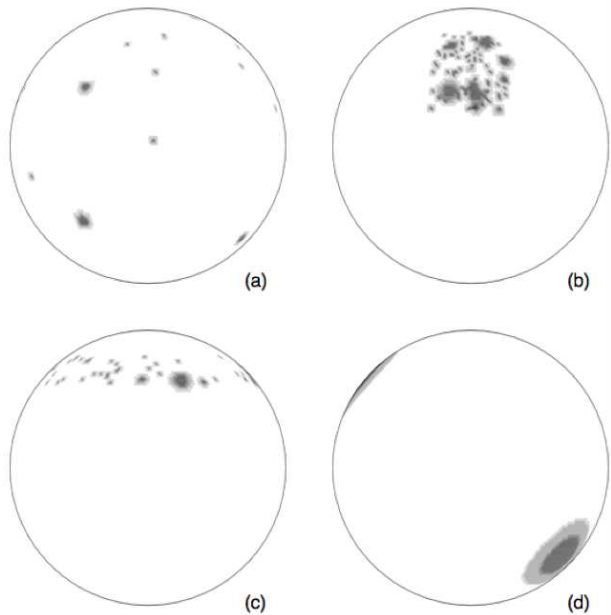} 
\caption{Spot configurations, (a): \textit{Rand}, random spots distributed over the entire stellar surface, (b): \textit{Vida1}, random spots at active longitude and latitude, (c): \textit{Granzer}, random spots at an active latitude range, as in \citet{Granzer2000}, (d): \textit{Vida2}, two large spots separated by 150\degree\ in longitude, similar to \citet{Vida2009}.}
\label{fig:spotmapsGVR}
\end{figure}

\begin{figure*}
\begin{minipage}{168mm}
 \includegraphics[width=84mm]{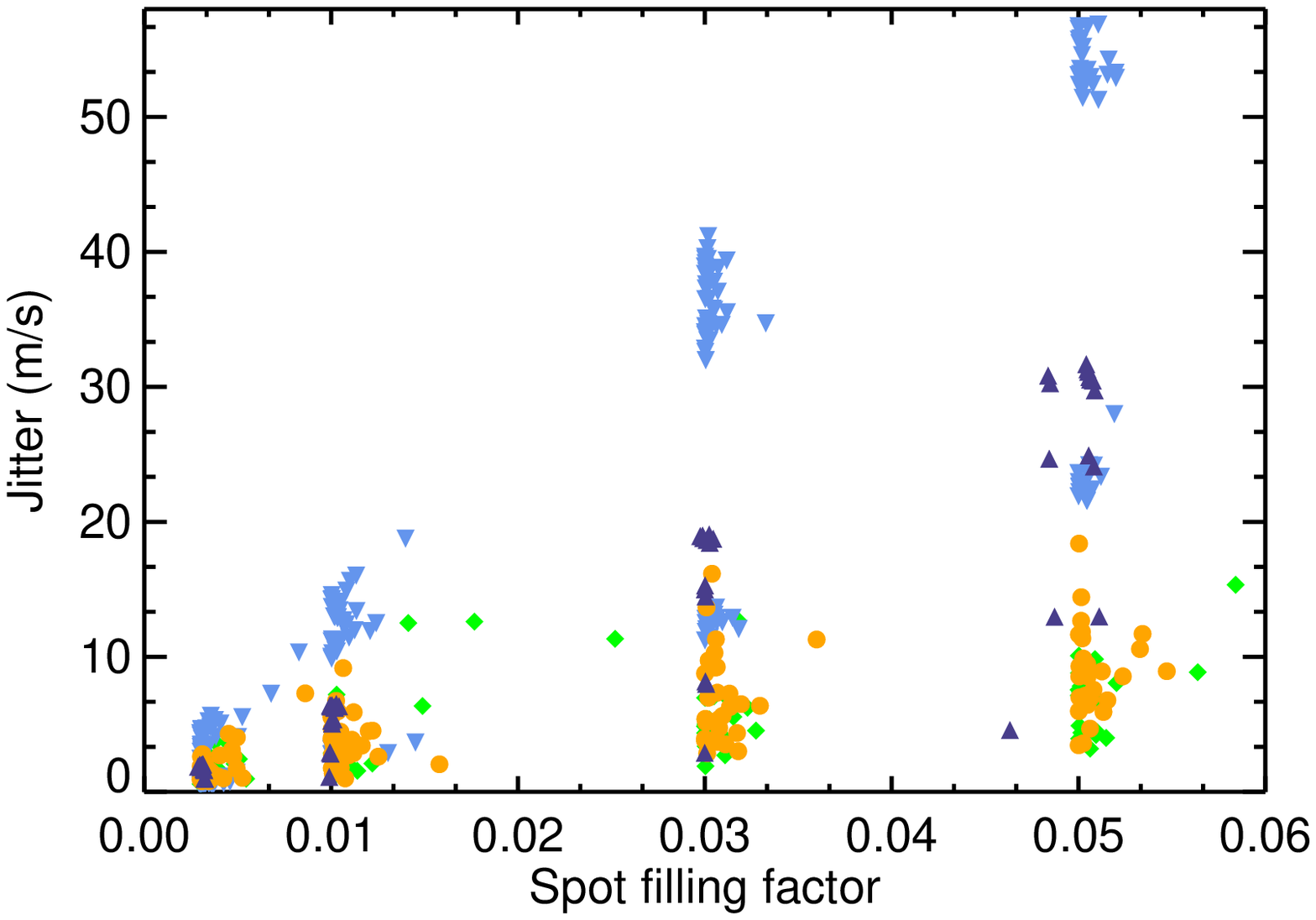} 
 \includegraphics[width=84mm]{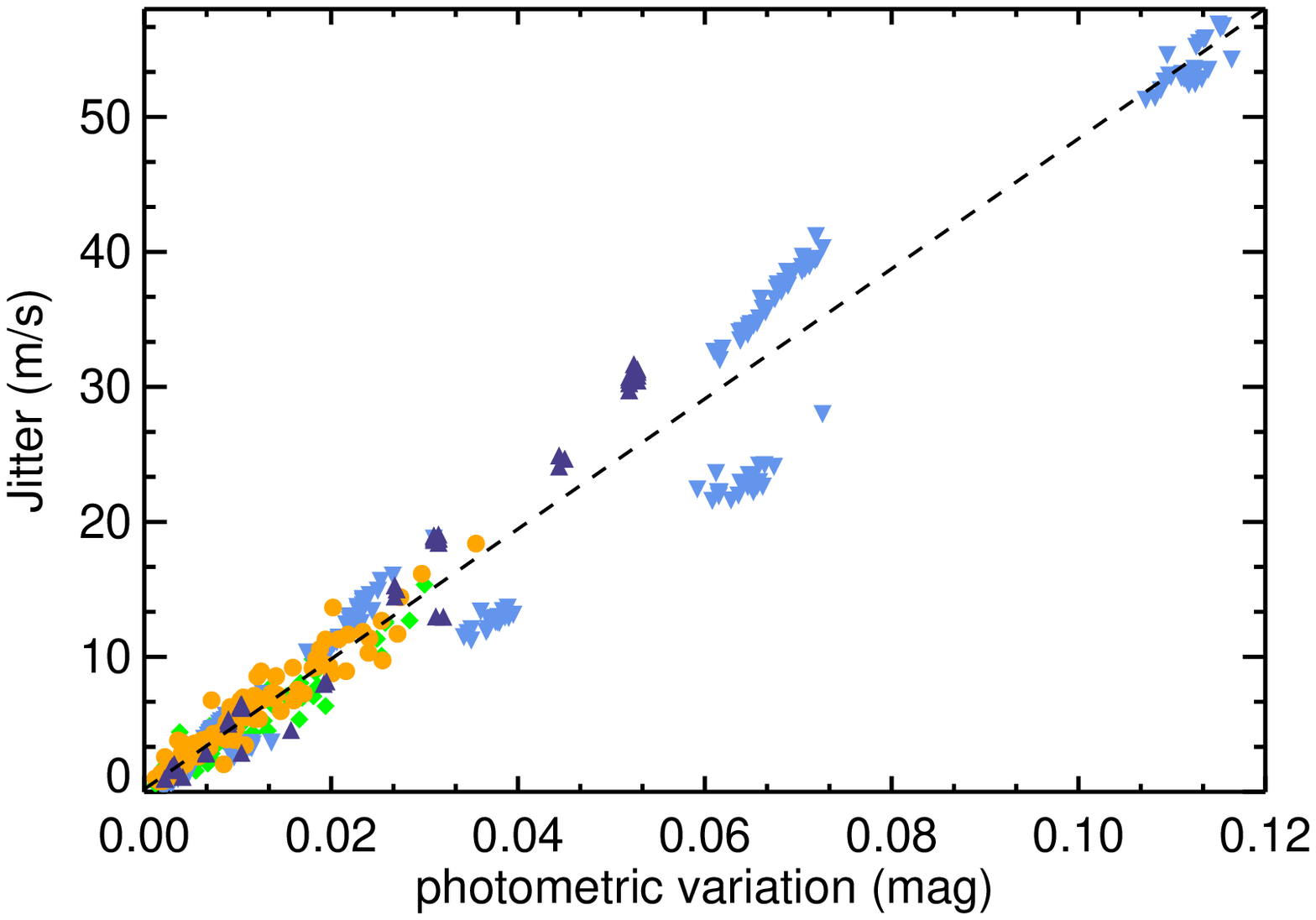} 
\caption{\textit{Left:} Jitter from spots as a function of filling factor for the four different spot configurations shown in Figure \ref{fig:spotmapsGVR}: random spots (green diamonds), spots at an active latitude (orange circles), spots in a distinct active region following \citet{Vida2009} (light blue, downward triangles, Vida1), and two distinct spots separated by 150 degrees in longitude, at differing latitudes (dark blue, upward triangles, Vida2). \textit{Right:} Jitter from the same spot configurations plotted as a function of V-band photometric variation of star. Whereas the jitter as a function of filling factor is highly dependent on spot configuration, when plotted as a function of photometric variation the resulting jitter shows almost no dependence on spot configuration but instead follows a linear relationship, which is shown by the overplotted dashed line (a linear fit to all data points). Vida1 configurations show a bit more deviation from the fit line and also often resulted in significantly higher jitter for the same filling factors compared with the other three distributions.}
\label{jitt_ff}
\end{minipage}
\end{figure*}

We created three different sets of spot configurations based on \cite{Granzer2000} and \cite{Vida2009}: \textit{Granzer}, with random spots only emerging at an active latitude range,  \textit{Vida1}, with small random spots clustered in a distinct active region of both longitude and latitude, and  \textit{Vida2}, with two large spots separated by 150\degree\ in longitude, but varying in latitude (see Figure \ref{fig:spotmapsGVR}).  We also created a set of random spot distributions with a range of filling factors.  We compare the resulting jitter from each configuration in Figure \ref{jitt_ff} (left).  The level of jitter, on average, increases with spot filling factor, however at a given filling factor the amount of jitter generated by a given spot map is highly dependent on the configuration of spots.  

Figure \ref{jitt_ff} (right) shows the RV jitter as a function of photometric variation of the star.  Regardless of spot configuration, the jitter increases linearly with photometric variation.  A line fit to all the data points is overplotted to illustrate this trend. Thus, stellar photometric variation is a much more useful predictor of RV jitter than spot filling factor.  This conclusion is supported by observational evidence from, e.g. \cite{Lanza2011}, who used high-precision optical photometry by the MOST satellite to map the longitudinal distribution of active regions in late-type stars and to predict the RV jitter caused by the spots. \cite{Aigrain2012} presented a method that predicted activity-induced jitter with high-precision time series photometry from a simple spot model, and tested the method using MOST and SOPHIE observations of the planet host HD189733.  Most recently, \cite{Cegla2014} used both Kepler light curves and GALEX data to examine the relation between RV jitter and photometric variation, concluding that for magnetically quiet stars the correlation was strong enough that photometric measurements can be used to directly estimate jitter.

\begin{figure}
 \includegraphics[width=84mm]{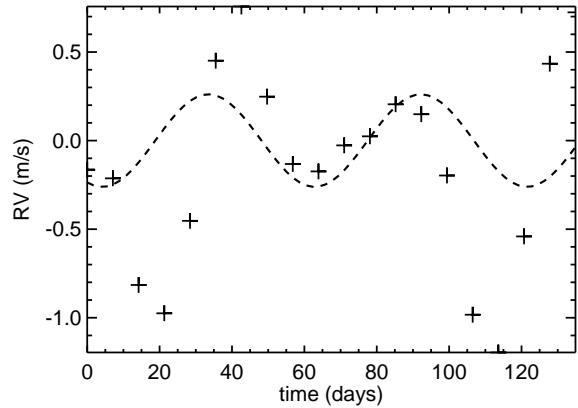} 
\caption{Simulated radial velocity from planet + noise + jitter.  20 ``observations" (crosses) have been evenly distributed over an ``observing run" with a length of 142 days.  The planetary contribution to RV signal is overplotted (dashed line), and generated from a simulated system of a 0.5M$_{\odot}$ star with a 1 M$_{\oplus}$ planet on a 58.7 day orbit. A single 3-degree spot on the stellar surface contributed to the jitter. The stellar rotation period is 7.7 days.  We can see that the RV jitter from the spot is significantly higher than the ``true" RV from the planet in many observations.   }
\label{RV_planet_plus_jitter}
\end{figure}

\subsection{RV from simulated planet}

We generated RV curves resulting from orbiting planets using Kepler's Third Law as is described in Paper I.  Figure \ref{RV_planet_plus_jitter} shows the sinusoidal curve resulting from an orbiting planet (dashed line), and the RV measurements that would result from this curve plus the added jitter of a 3-degree spot on the stellar surface (crosses).  
We investigated some  ``detection limits" for orbiting planets by using a Lomb-Scargle Periodogram and attempting to recover the period of the planet from our radial velocity curves.  These are not robust limits on planetary detection (see the following section) but simply a quick mechanism for analyzing the RV curves generated from the planet + spots model and how the noise introduced by the spots can obstruct planet detection.  For each spot configuration + simulated orbiting planet, we investigated a number of different, evenly spaced sets of ``measurements" to obtain an RV curve.  We then created a Lomb-Scargle Periodogram of each RV curve and identified the planetary period in the resulting peaks.  If there was a peak corresponding to the planetary period, we calculated the False Alarm Probability (FAP) of that period based on its Scargle Power.  We calculated the FAP by randomly rearranging the RV data points $N(N-1)$ times (where $N$ is a the number of data points) and then plotting the randomized RV array against the time array, and taking a LS-Periodogram of the resulting curve.  We then found any peaks in the periodogram, and their Scargle power, and compared these values to the Scargle power of the original period detection.  The FAP is the fraction of trials in which we found peaks with Scargle power higher than that of the peak corresponding to the actual period in the original, non-randomized data.  This method of calculating the FAP is useful because it incorporates the uncertainties already in the data, and does not require a gaussian noise distribution.

We consider a FAP of $\leq$ 1\% to be a ``detection" and anything higher than this to be a non detection.  We investigated three contrast ratios,  low, $\Delta T = 100$, medium, $\Delta T = 500$, and high, $\Delta T = 900$, over a range of filling factors, 0.1\% (Solar) 0.3\% (Solar maximum), 1\%, 3\%, 10\%, 30\%, 50\%, and 80\%, with a ``random" spot distribution (see Figure \ref{fig:spotmapsGVR}, a).  We simulated four different planets around each star with masses M = 1\,$M_{\oplus}$ (Earth), 5\,$M_{\oplus}$ \& 10\,$M_{\oplus}$ (Super-Earths), and 20\,$M_{\oplus}$ (Neptune).  We also investigated two photospheric temperatures, $T_\mathrm{eff} = 3150$ and $T_{eff} = 3650$, corresponding to stellar masses $M_{*} = 0.15\,M_{\odot}$ and $M_{*} = 0.50\,M_{\odot}$, respectively.  For each stellar temperature, we used a semi-major axis of orbit for the orbiting planets within the classically defined conservative HZ for the star, from \citet[][see Section \ref{sec:Habitability}]{Kopparapu2013}.  These orbits correspond to periods of 21.3 days and 84.7 days for $T_\mathrm{eff} = 3150$ and $T_\mathrm{eff} = 3650$, respectively.  Unless otherwise specified, we used a default stellar rotation period of 7.7 days.

\begin{figure*}
\begin{minipage}{168mm}
 \includegraphics[width=84mm]{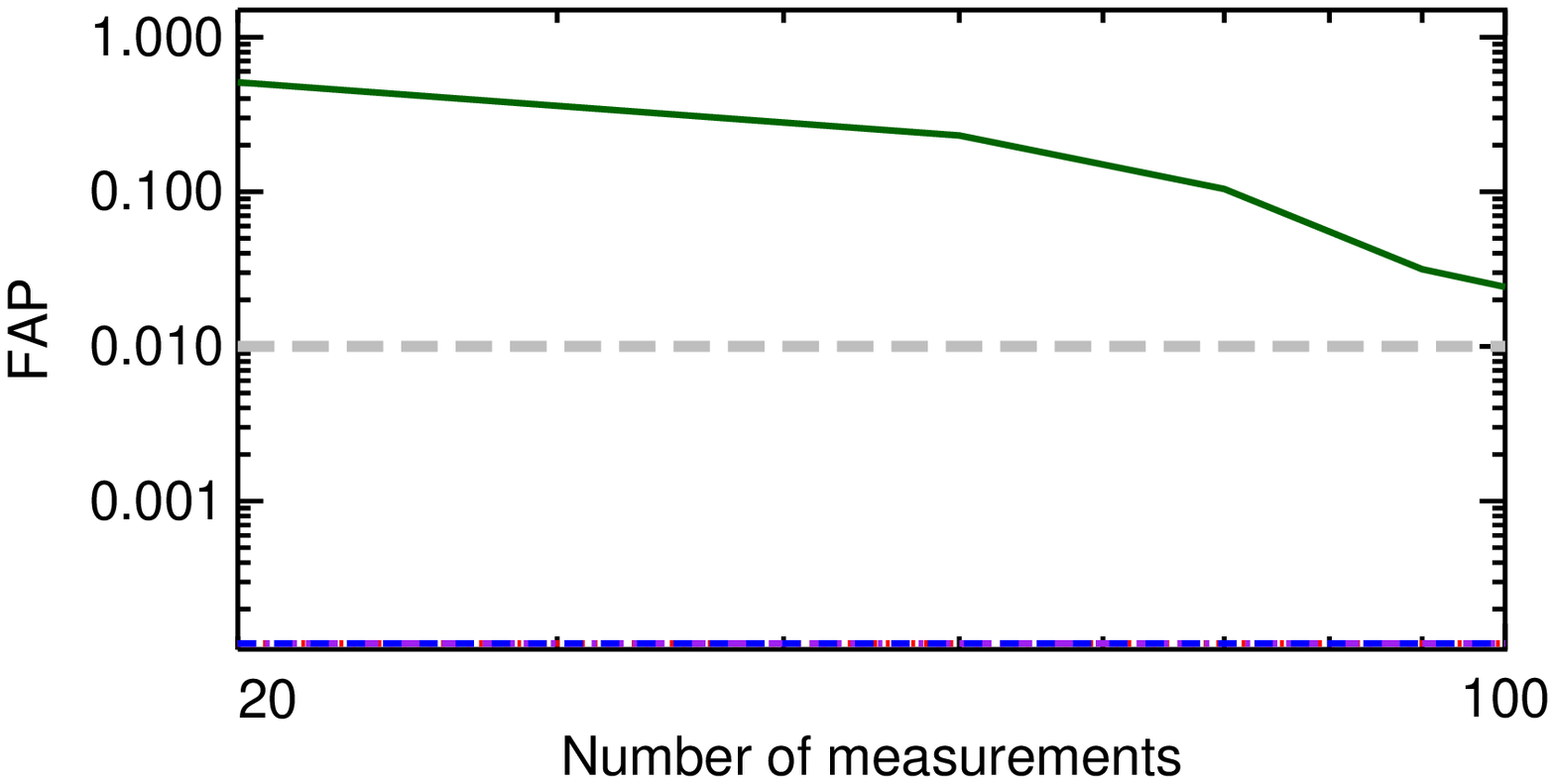} 
  \includegraphics[width=84mm]{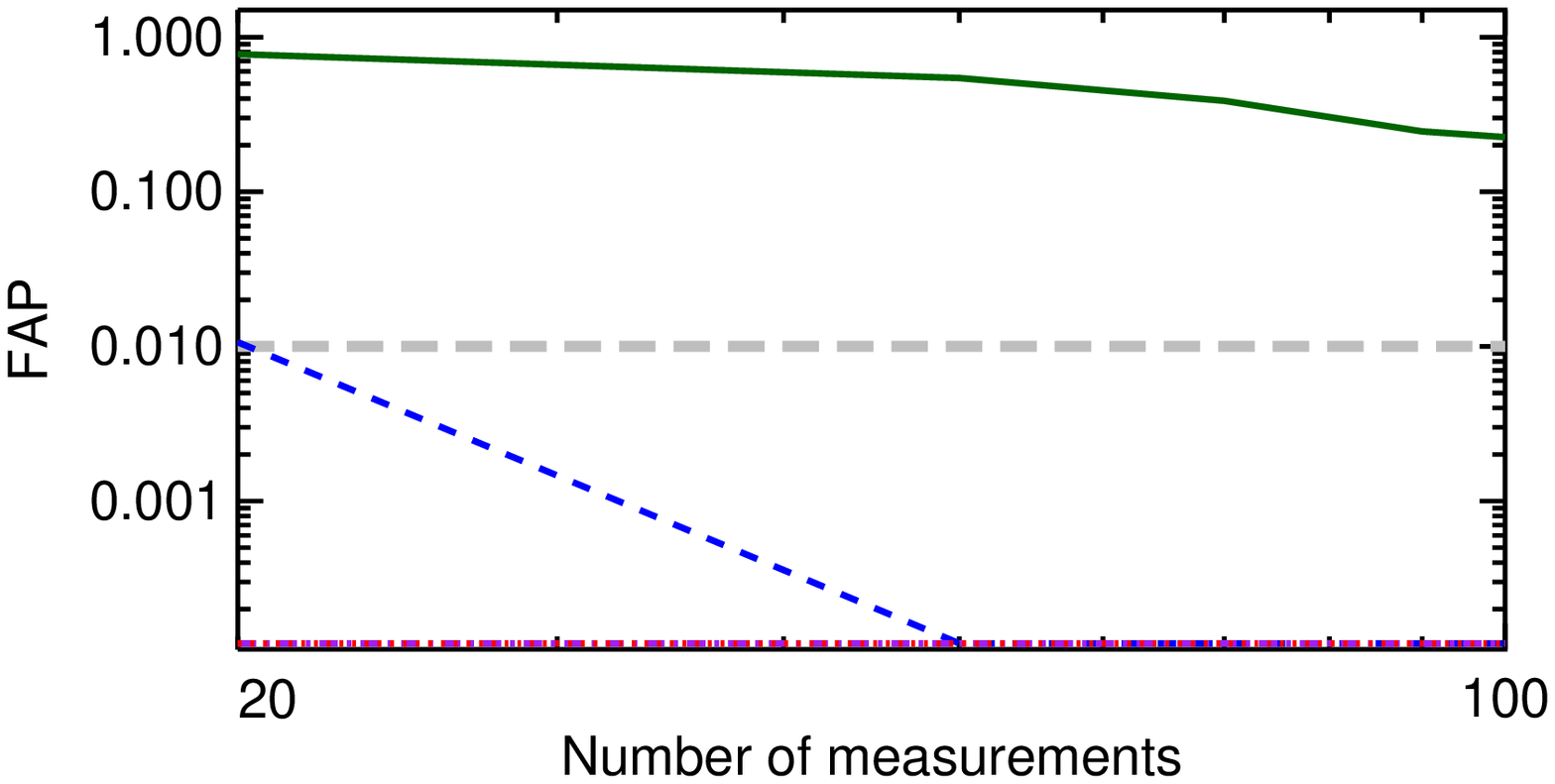} 
\caption{Lomb Scargle (LS) false alarm probability of planet detection as a function of number of measurements made during a 112-day period. The two Solar filling factor cases: 0.1\% (left) and 0.3\% (right) are shown.  Stellar temperature is $T_\mathrm{eff} = 3650$ and stellar mass is $M_{*} = 0.15\,M_{\odot}$.  Spot temperature contrast is $\Delta T = 500$, for s spot temperature of $T_\mathrm{spot} = 3150$.  Planets are on an 84.7-day orbit which corresponds to an orbital radius within the HZ for this star. Four planet masses are included: M = 1\,$M_{\oplus}$ (solid green line), 5\,$M_{\oplus}$ (dashed blue line), 10\,$M_{\oplus}$ (dot-dashed purple line), and 20\,$M_{\oplus}$ (dotted red line). The M = 1\,$M_{\oplus}$ remains above the detectability (defined as a FAP of 1\% or less) threshold (indicated by the dashed grey line) for all numbers of measurements. The 5\,$M_{\oplus}$ planet is also detectable in all instances, although at 20 observations the FAP just grazes the detectability limit in the 0.3\% filling factor case (right).  10\,$M_{\oplus}$ and 20\,$M_{\oplus}$ planets have a FAP of zero consistently in both cases. Note that zero is plotted at a line just above the x-axis.}
\label{fig:ffplot_Solar}
\end{minipage}
\end{figure*}

Figure \ref{fig:ffplot_Solar} shows the FAP as a function of the number of measurements (``observations") taken over a 112-day observing run for the Solar filling factor cases of the $T_{eff} = 3650$ star with medium contrast ($T_{spot} = 3150$, $\Delta T = 500$).  The solid green line represents the M = 1\,$M_{\oplus}$ planet and is not detectable even with 100 observations.  The other planets are all detectable at these filling factors.  

\begin{figure*}
\begin{minipage}{168mm}
 \includegraphics[width=84mm]{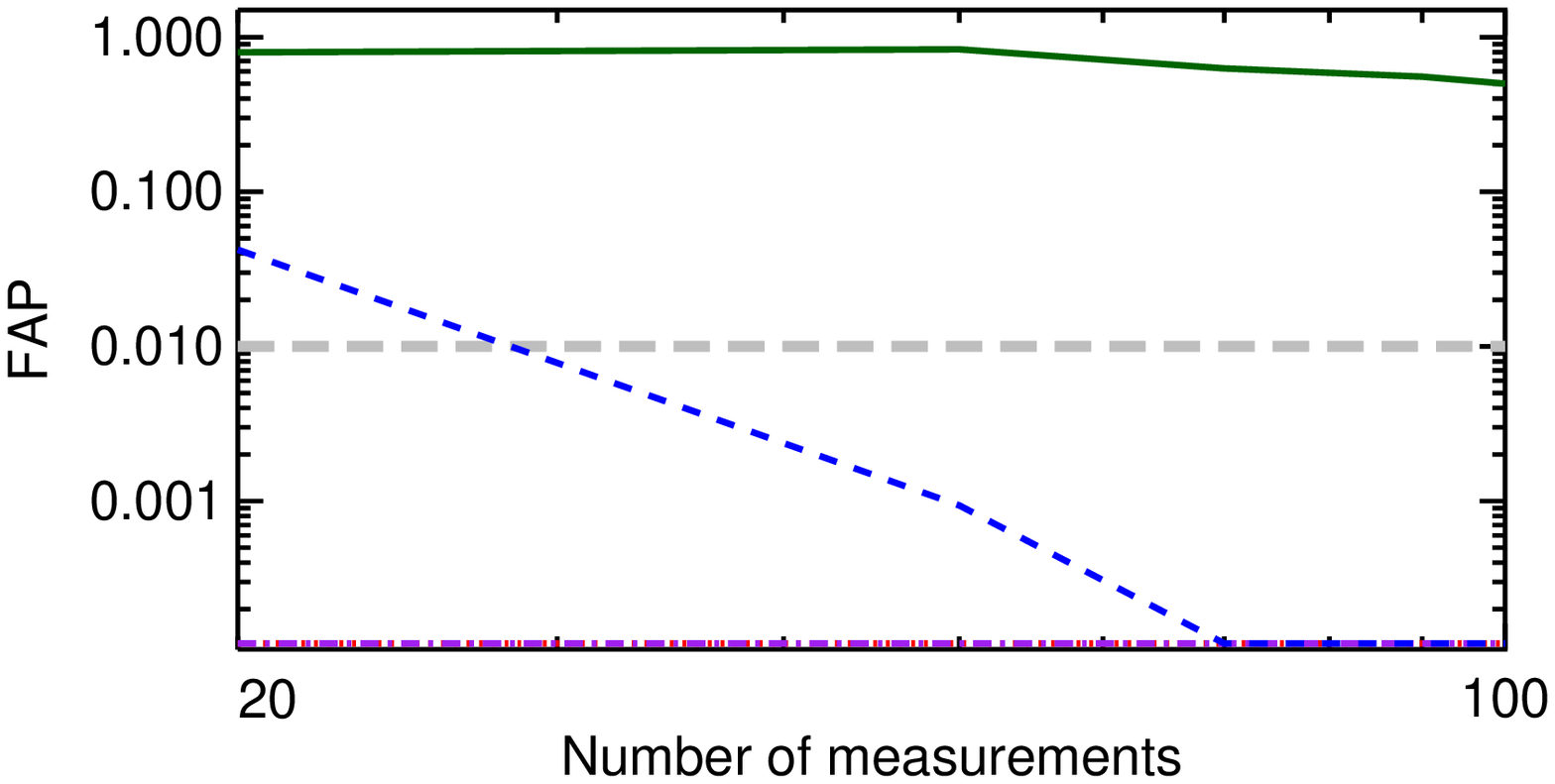} 
  \includegraphics[width=84mm]{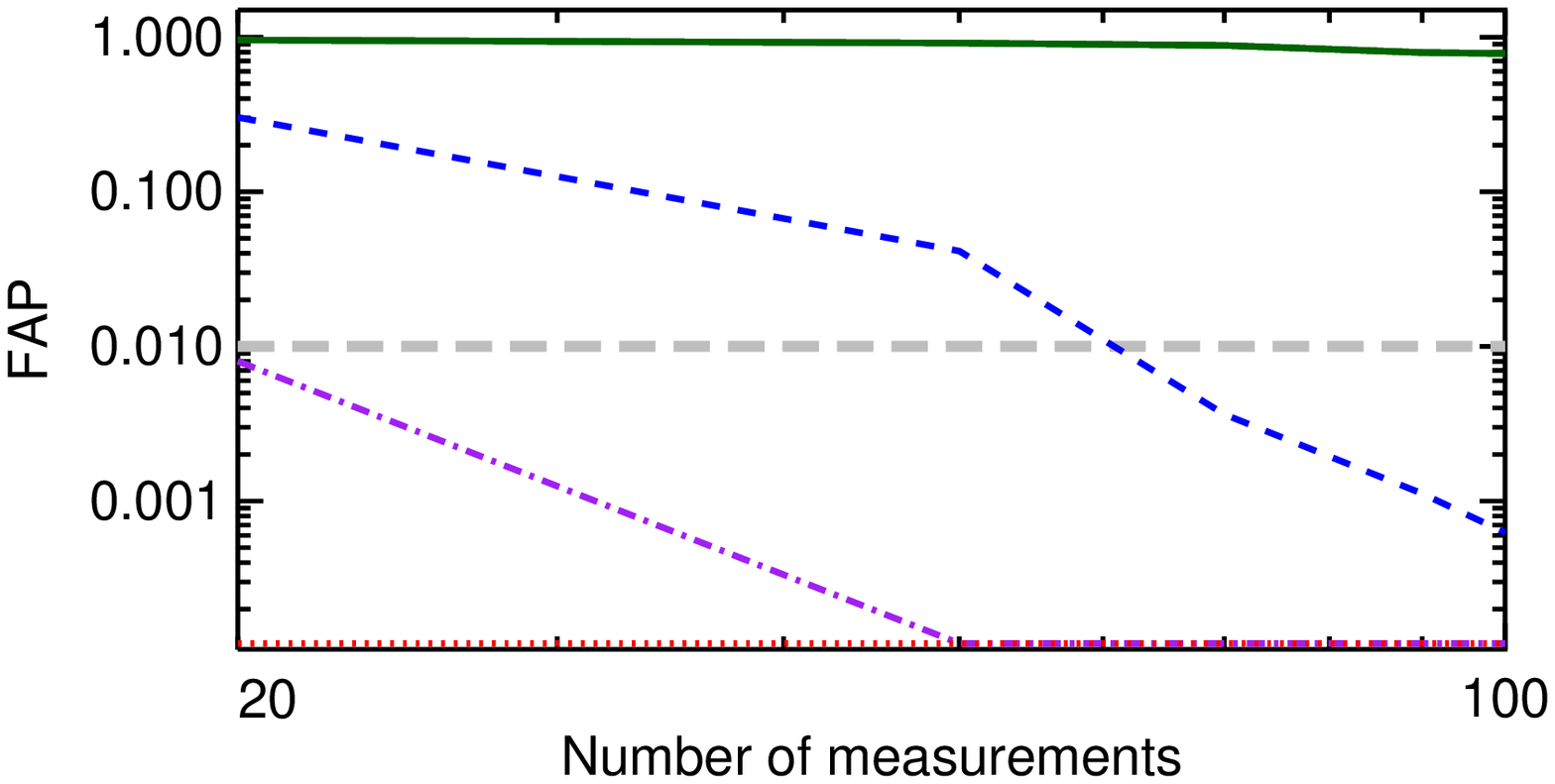} 
 \includegraphics[width=84mm]{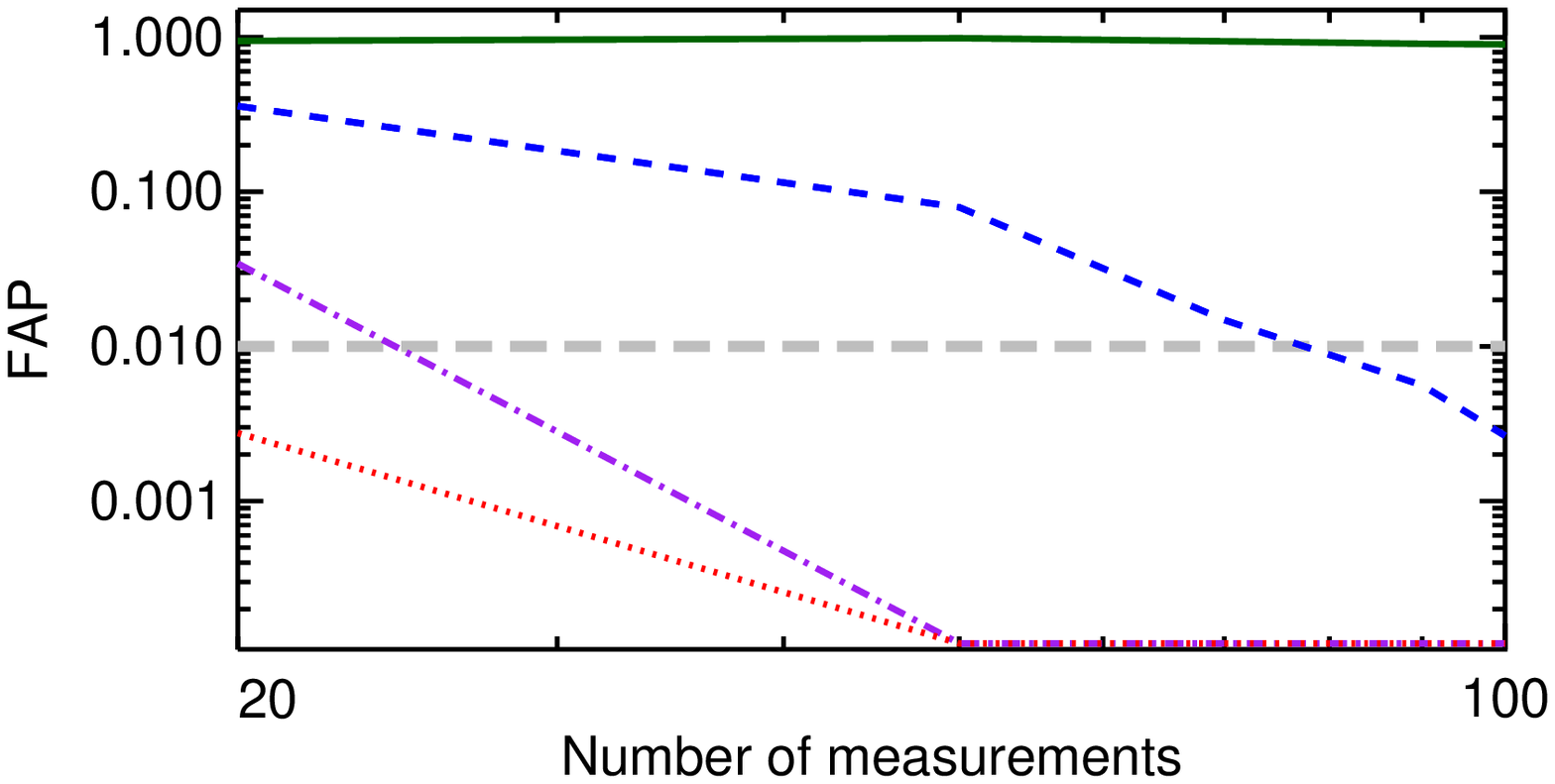} 
  \includegraphics[width=84mm]{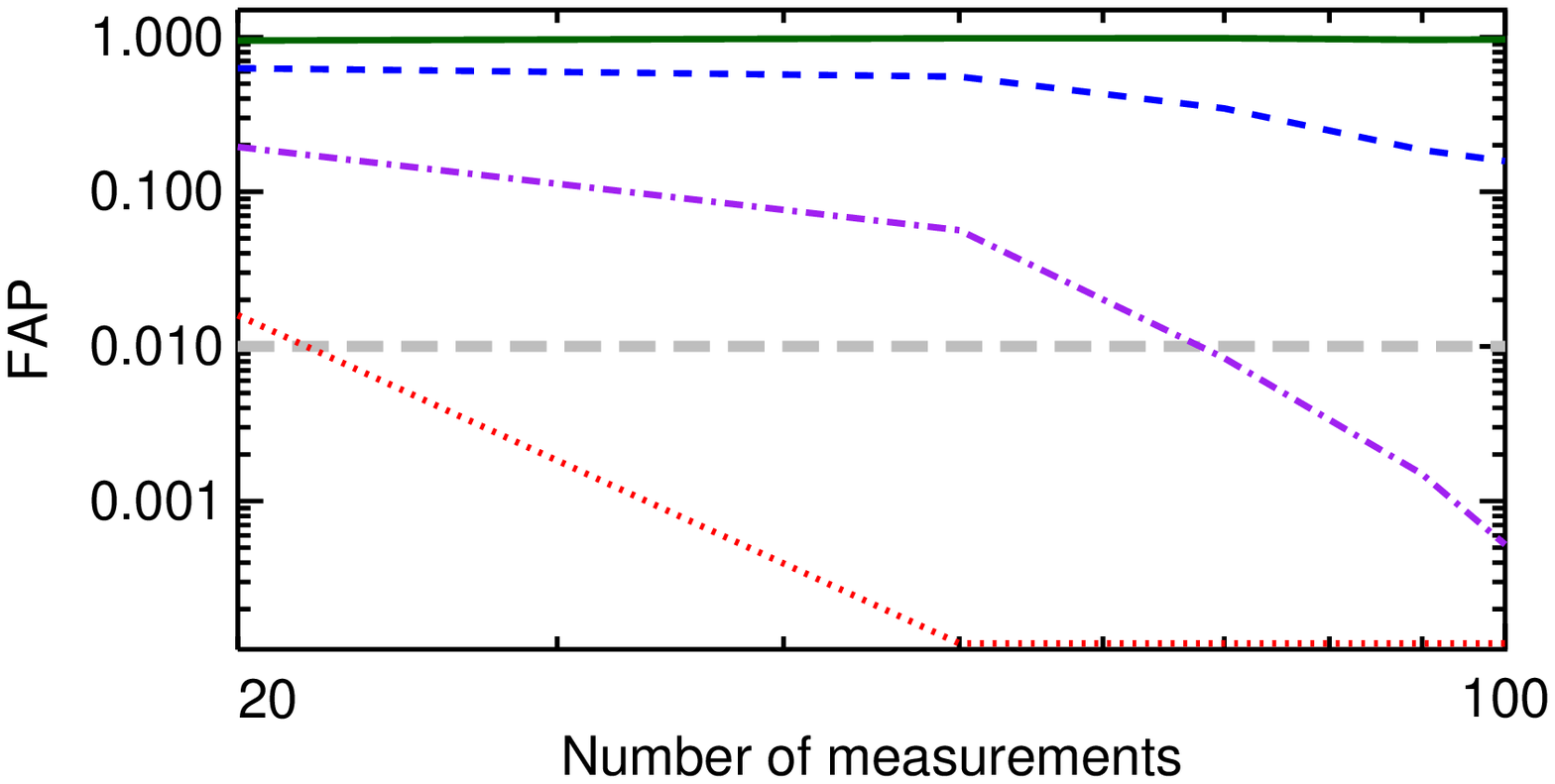} 
\caption{LS FAP of planet detection as a function of number of measurements for the same star and planet cases as in Figure \ref{fig:ffplot_Solar}.  Four ``intermediate" filling factor cases are illustrated: 1\% (top, left), 3\% (top, right), 10\% (bottom, left) and 30\% (bottom, right).  Increasing the filling factor from 1\% to 10\% (left) requires an increase from 30 to 80 measurements for the 5\,$M_{\oplus}$ planet to be detected.  Grey dashed line indicates FAP detection threshold:  planets are considered ``detectable" where FAP falls below this line.  The same trend is seen with the 10\,$M_{\oplus}$ planet (right) where an increase from 3\% filling factor to 30\% requires a jump from 20 to 70 measurements to reach the detectability threshold.}
\label{fig:ffplot_intermediate}
\end{minipage}
\end{figure*}

``Intermediate" filling factors, 1\% - 30\%, are shown in Figure \ref{fig:ffplot_intermediate}.  As the filling factor increases, more planets evade detectability to a higher number of observations.  At 1\% filling factor, the 1\,$M_{\oplus}$ and 5~$M_{\oplus}$ are both undetectable with 20 observations, but the 5~$M_{\oplus}$ planet reaches a sufficiently low FAP of detectability at around 30 observations.  By 30\% filling factor, all of the planets are initially undetectable, although the 20\,$M_{\oplus}$ planet reaches detectability with just a slight increase in number of observations, and the 10\,$M_{\oplus}$ planet is detectable in 70 observations.  However, both the 1\,$M_{\oplus}$ and 5~$M_{\oplus}$ planet remain undetectable even up to 100 observations.  

\begin{figure*}
\begin{minipage}{168mm}
 \includegraphics[width=84mm]{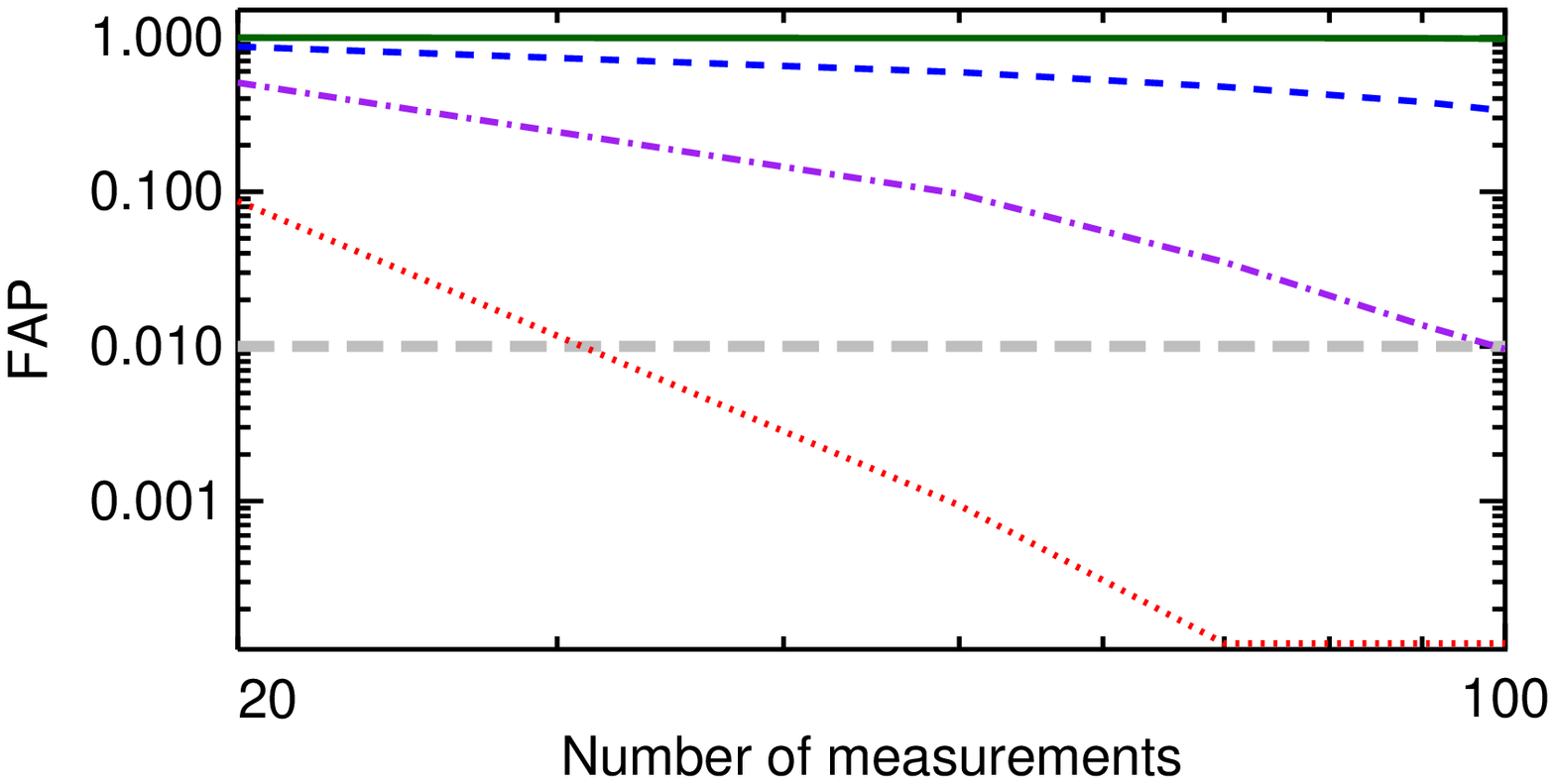} 
  \includegraphics[width=84mm]{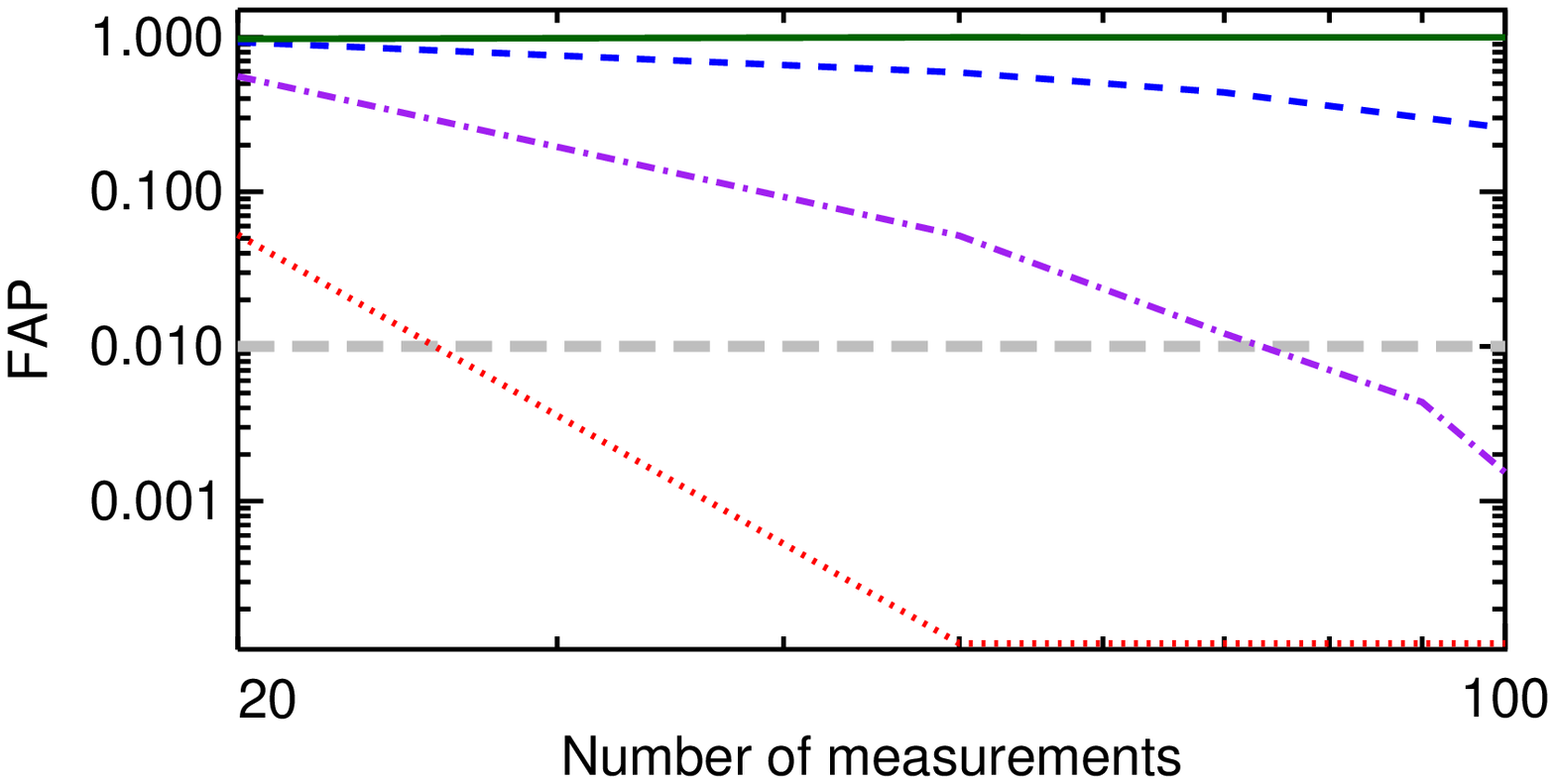} 
\caption{LS FAP of planet detection as a function of number of measurements for the same star and planet cases as in Figure \ref{fig:ffplot_Solar}. ``High" filling factors of 50\% (left) and 80\% (right) are illustrated. The number of observations required to achieve detectability for all planets decreases slightly in the 80\% filling factor case from the 50\% case. The 80\% case more closely resembles the 30\% case from Figure \ref{fig:ffplot_intermediate}. This trend was expected since with spot filling factors of above 50\% the spots dominate the stellar surface and the level of spot-induced jitter actually decreases. }
\label{fig:ffplot_high}
\end{minipage}
\end{figure*}

``High" filling factors, 50\% and 80\%, are shown in Figure \ref{fig:ffplot_high}.  Detectability is the lowest for all planets in the 50\% filling factor case, and actually improves slightly at 80\% filling factor. The 80\% filling factor case more closely resembles the 30\% filling factor case shown in Figure \ref{fig:ffplot_intermediate}.  This makes sense since a star with an  80\% coverage of dark spots is essentially the same as a dark star with 20\% coverage of bright spots (see Figure \ref{Vmag_ff}).

\begin{figure*}
\begin{minipage}{168mm}
 \includegraphics[width=84mm]{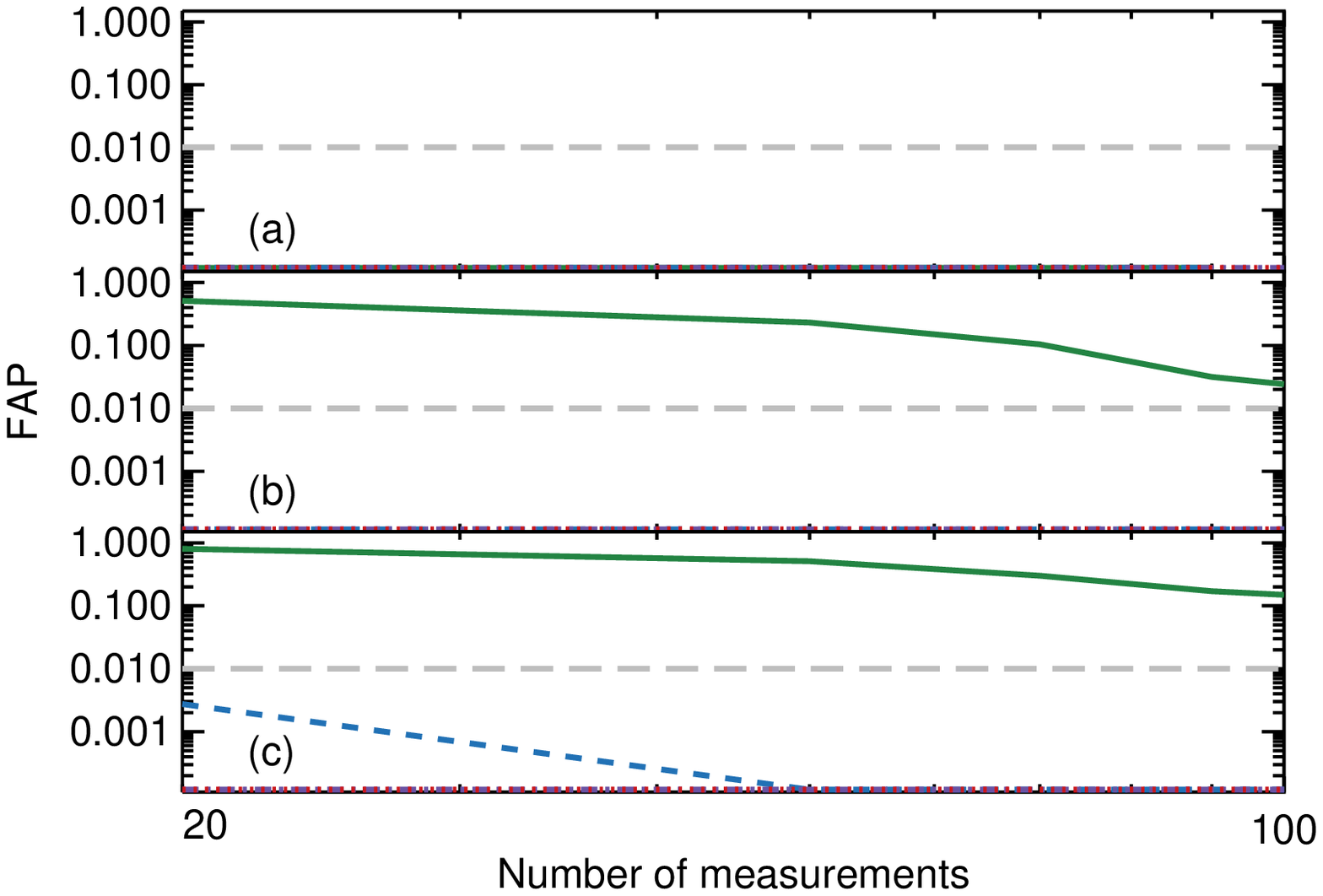} 
  \includegraphics[width=84mm]{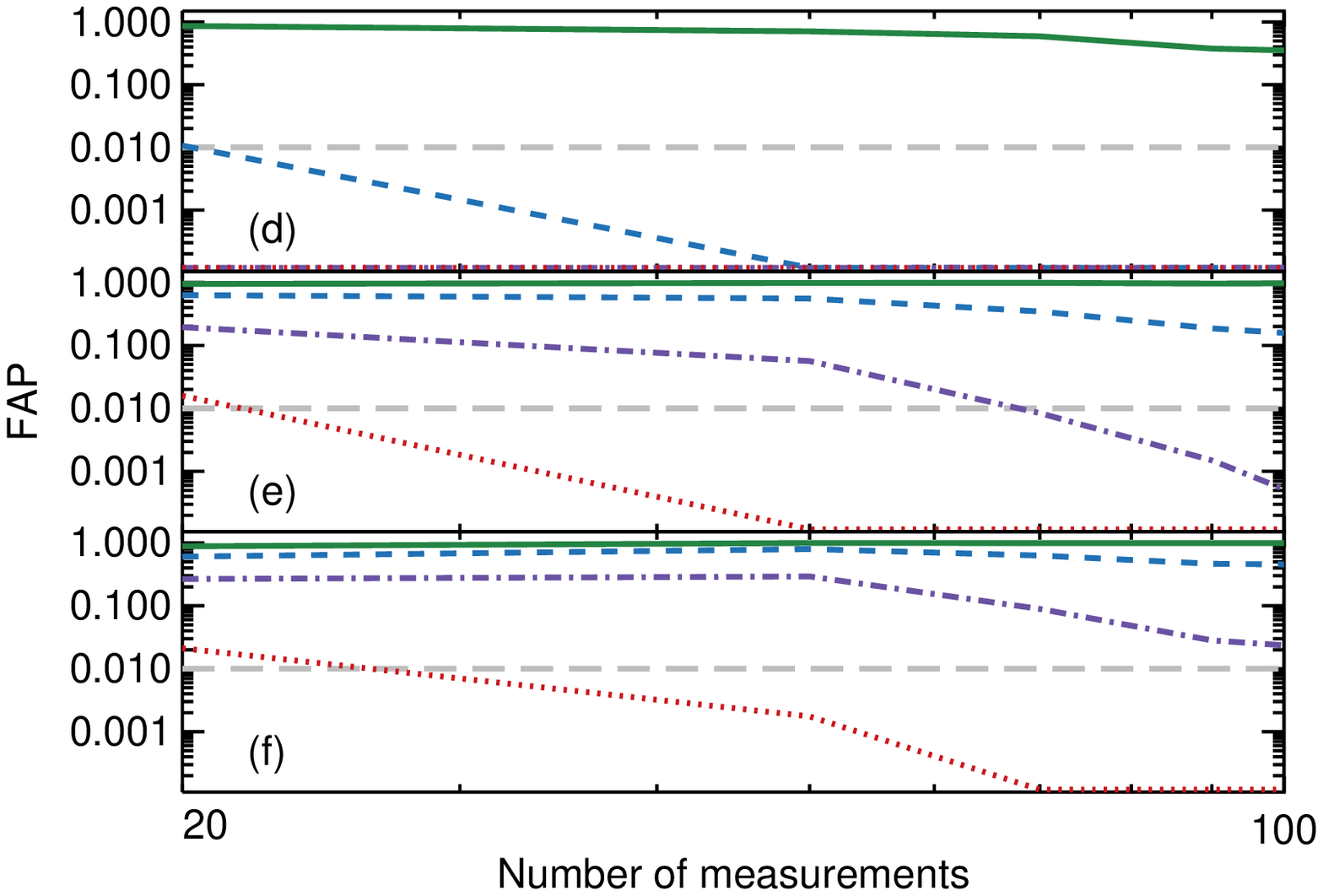} 
\caption{LS FAP of planet detection as a function of number of measurements for a $T_\mathrm{eff} = 3650$ star. Solar filling factor (0.1\%) case is shown on the left at low (a) medium (b) and high (c) temperature contrasts, and 30\% filling factor is shown on the right, also at low (d), medium (e), and high (f) temperature contrasts.  1\,$M_{\oplus}$ planet (solid green line) is only detectable in low filling factor, low contrast case (a).  Note that in the (a) panel of the left plot, FAP is 0 for all planet masses.} 
\label{fig:ffplot_3contrasts_3650}
\end{minipage}
\end{figure*}

\begin{figure*}
\begin{minipage}{168mm}
 \includegraphics[width=84mm]{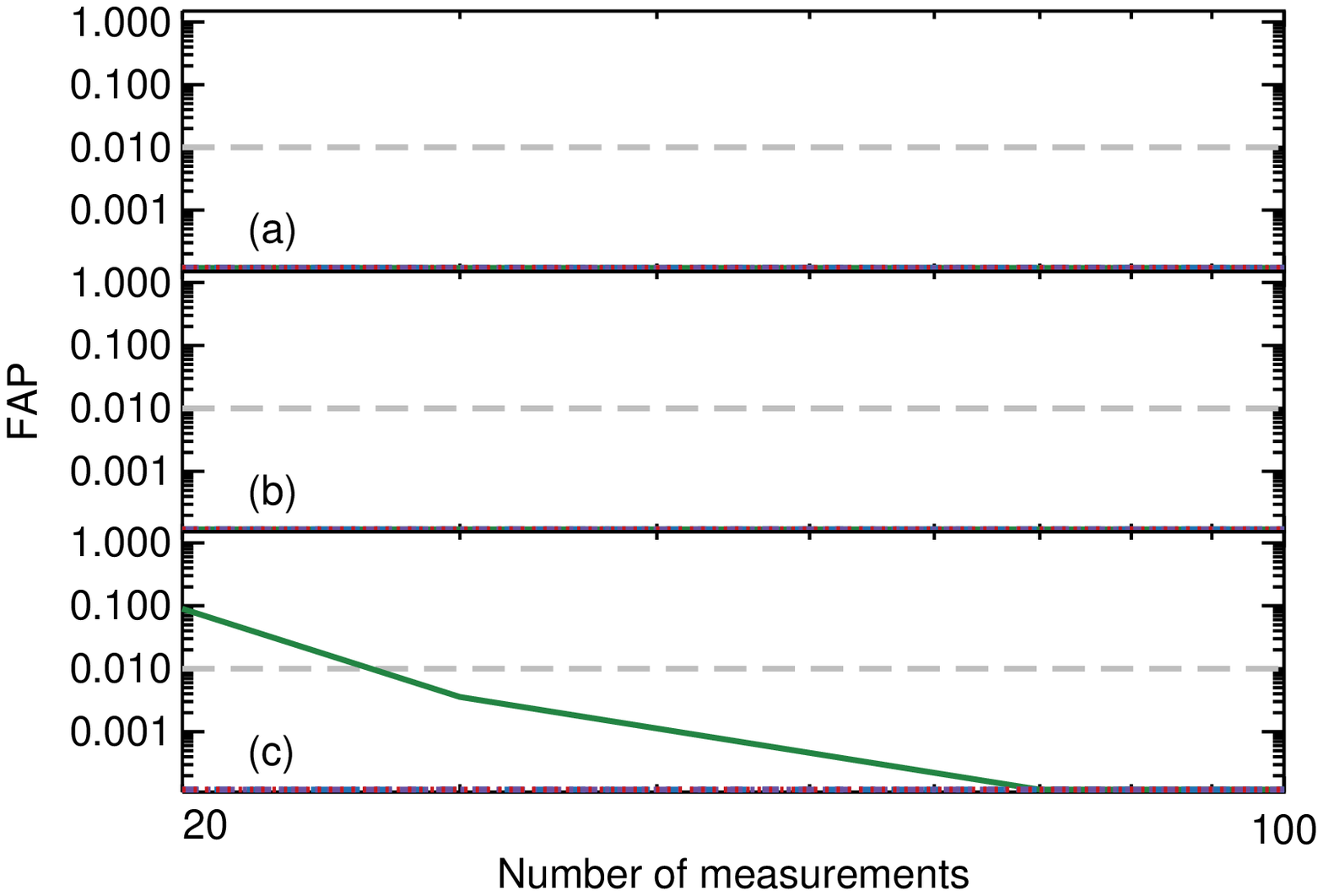} 
  \includegraphics[width=84mm]{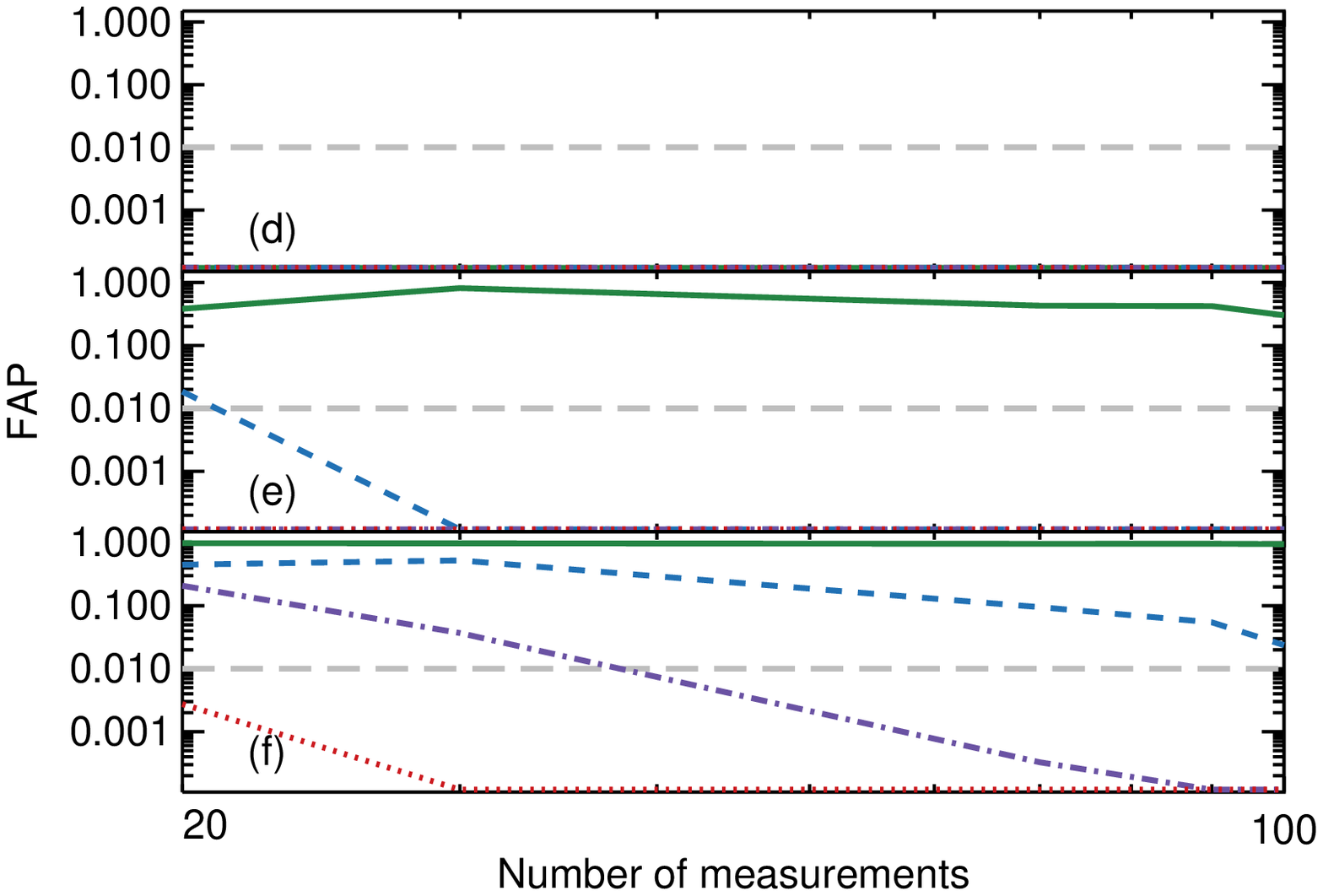} 
\caption{LS FAP of planet detection as a function of number of measurements for the same filling factor and temperature contrast cases as in Figure \ref{fig:ffplot_3contrasts_3650}, but for a $T_\mathrm{eff} = 3150$ star. In panels (a), (b), and (d) FAP is 0 for all planets.}
\label{fig:ffplot_3contrasts_3150}
\end{minipage}
\end{figure*}

The effect of temperature contrast on detectability is shown in Figures \ref{fig:ffplot_3contrasts_3650} and \ref{fig:ffplot_3contrasts_3150}.  Figure \ref{fig:ffplot_3contrasts_3650} illustrates the $T_\mathrm{eff} = 3650$ ($M_{*} = 0.50\,M_{\odot}$) case.  The Solar filling factor (0.1\%) case is shown on the left with low (a), medium (b), and high (c) contrast ratios.  For the low contrast case, all planets are detectable.  The  1\,$M_{\oplus}$ planet is not detectable for medium or high contrast at this filling factor.  The plot on the right in Figure \ref{fig:ffplot_3contrasts_3650} shows the 30\% filling factor case, again at low (d), medium (e), and high (f) contrasts.  Increasing contrast inhibits detectability of all planets, requiring more observations to yield a detection.  At high contrast, only the 20\,$M_{\oplus}$ planet achieves detectability in this observational scenario.  Figure \ref{fig:ffplot_3contrasts_3150} shows the same cases, but for the $T_\mathrm{eff} = 3150$ ($M = 0.15\,M_{\odot}$) star. A similar trend is observed, although detectability at a given planet mass + contrast + filling factor case is significantly better around this star, due to the combination of smaller stellar mass and closer-in HZ orbit.   Since we used the same observing period of 112 days for both stars, the data covers more orbits for the smaller star, since the closer-in HZ means a shorter orbital period.  When the data are phase folded, more of the activity noise can be averaged out because at a given planetary phase there are more data points.  We repeated the experiment using an observing period of 445 days for the 3650\,K M dwarf (with a HZ orbit of 84.7 days), which corresponds to the same planetary orbital phase coverage as the 112--day observing period of the 3150 K star (HZ orbit of 21.3 days).  In these observations the detection limits were more similar to the 3150\,K case due to the similar phase coverage, although still not identical because the stellar RV generated by a planet of a given mass is different in the two systems due to the different stellar masses and HZ orbits corresponding to the different stellar temperatures. 

At a given filling factor, the spot contrast makes a huge difference in the number of observations required to detect these HZ planets.  Thus, better constraints on spot temperatures would help to better plan observing strategies when searching for HZ planets, especially around active stars.  Even at low filling factors, the difference in spot temperature could mean the difference between finding a HZ planet and missing it.  In Figure \ref{fig:ffplot_3contrasts_3650}a (low contrast case) the 1\,$M_{\oplus}$ planet would be detected even with a low number of observations.  However, when the contrast is increased (\ref{fig:ffplot_3contrasts_3650}b), the FAP for this planet is still just above the detection threshold, even at 100 observations.  When the contrasts is highest (\ref{fig:ffplot_3contrasts_3650}c), the FAP is an order of magnitude over the detection limit with 100 observations, and would require significantly more observations to achieve detectability.  We extended the observations of the 1\,$M_{\oplus}$ planet (orbiting a 3650\,K star with a 0.1\% spot filling factor), and found that even at 500 observations the FAP still did not drop below the detection threshold of 0.01.

\begin{figure}
 \includegraphics[width=84mm]{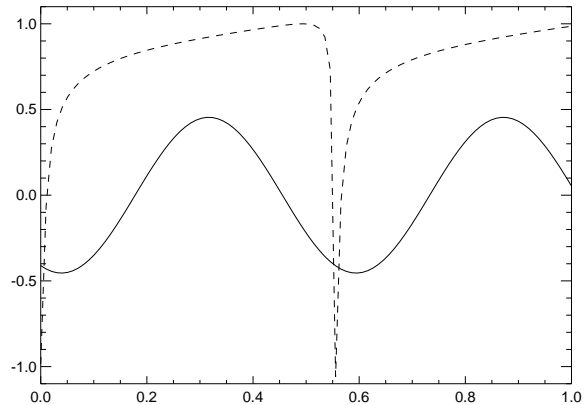}  
\caption{Stellar radial velocity curves generated by orbiting planets with identical parameters but different orbital eccentricities: 0.0 (solid line) and 0.9 (dashed line). Y-axis is in arbitrary velocity units where the high eccentricity case has been normalized to 1. X-axis is also normalized to one in arbitrary time units. }
\label{fig:example_ecc}
\end{figure}

\begin{figure}
 \includegraphics[width=84mm]{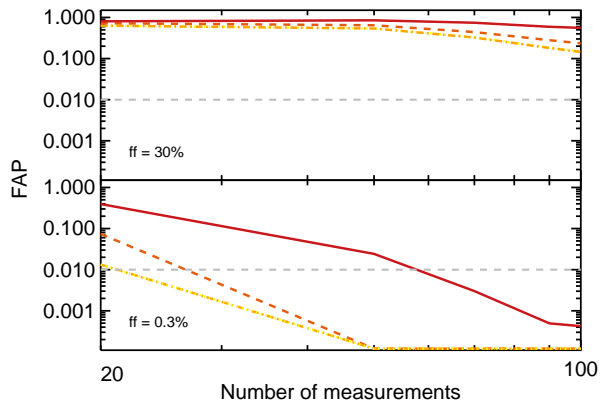}  
\caption{LS FAP of planet detection as a function of number of measurements for a 5\,$M_{\oplus}$ planet with four different orbital eccentricities: 0.0 (red solid line), 0.3 (dark orange, dashed line), 0.5 (orange dash-dot line), 0.8 (yellow dotted line).  FAP detected threshold is indicated by dashed grey line.  All orbits have the same period, 84.7 days. The stellar temperature is 3650\,K and the spot temperature is 3150\,K ($\Delta T = 500\,K$).  Spot filling factors of 30\% (top) and 0.3\% (bottom) are compared.  The same trend is seen in both: high eccentricities have lower FAP.  The 0.5 and 0.8 eccentricities, however, are almost exactly the same. }
\label{fig:ffplot_ecc}
\end{figure}

We also tested the effect of orbital eccentricity by investigating a 5\,$M_{\oplus}$ planet with four orbital eccentricities ranging from 0.0 to 0.8. Figure \ref{fig:ffplot_ecc} illustrates a high (30\%) and low (0.3\%) filling factor case, and the same trend is seen in both: higher eccentricity causes a lower FAP (though the two highest eccentricity cases, 0.5 and 0.8, resulted in almost the exact same FAP for all measurements).  Thus, a planet with a more circular orbit would be more difficult to detect than a planet with higher eccentricity if all other stellar and planet parameters were equal, and if the signals are well-sampled in phase. This is due to the fact that a planet in a highly eccentric orbit will create an RV signal with a higher semi-amplitude, as seen in Figure \ref{fig:example_ecc}.  However, if the signal is not well-sampled, the signature of a high-eccentricity planet could be missed completely, because the RV curve is mostly relatively flat, with short-lived, high-amplitude peaks (or troughs). 

\section{uncertainty in derived planetary system parameters}

We hesitate to declare hard ``limits" on exoplanet detection due to the ever-increasing precision with which RV curves can be measured and the many techniques that are being constantly developed to improve RV data and remove spurious periodic signals in order to uncover planetary signatures, such as Fourier component analysis via pre-whitening, and local trend filtering (see, for example, \cite{Boisse2011a}, who corrected for activity noise by fitting sinusoids at the stellar rotation period and its two-first harmonics and were able to remove about 90\% of the RV jitter amplitude; and \cite{Hatzes2013}, who demonstrated that it is possible to extract the RV signal of a short-period Earth-mass planet despite stellar activity noise, if high-cadence observations are available). Developing technology is allowing spectra to be measured to such precision that the limit on derived radial velocities now stems from limits due to properties of the star itself.  \cite{Dumusque2012} analyzed high-precision radial velocities for $\alpha$ Centauri B, and identified the following contributions to the radial velocity signal: instrumental noise, stellar oscillation modes, granulation at the surface of the star, rotational activity, long-term activity induced by a magnetic cycle, the orbital motion of the binary system, light contamination from $\alpha$ Centauri A, and imprecise stellar coordinates.  They then attempted to model and remove each signal separately.  They extracted the RV signature of a small, approximately Earth-mass planet.  However,  \cite{Hatzes2013} reanalyzed the data using different methods to remove the stellar activity signal from the RV data, and was unable to recover a significant RV signal corresponding to the \cite{Dumusque2012} planet.  

Thus, we do not attempt to set ``limits" on radial velocity measurements based on this study alone.  Rather, in this work we are interested in the potential uncertainty in planetary system parameters derived from RV observations that can be introduced via contributed stellar jitter.  

To derive planetary parameters from our synthetic RV curves, we used the \cite{Eastman2013} EXOFAST suite of IDL routines.  EXOFAST was created to fit exoplanetary transits and radial velocity variations either simultaneously or separately, and characterize the parameter uncertainties and covariances using a Differential Evolution Markov Chain Monte Carlo method.  Since we are interested in RV jitter, we only fit the RV variations with EXOFAST.  Simultaneous fitting of transits and RV variations will be investigated in a later work.  

Markov Chain Monte Carlo (MCMC) algorithms are commonly used to fit data points to a model with multiple parameters.  EXOFAST uses the Metropolis-Hastings MCMC algorithm to sample the probability that a given model $M$ is correct, given the data, $D$: $P(M\textpipe D)$, starting with a trial set of model parameters (see below for a discussion of the specific parameterization we used for our models) and evaluating $\chi^{2}$ with respect to the data.  EXOFAST then calculates the ratio of the likelihood of a new set of parameters relative to the initial set.  Then a uniform random number is drawn from [0,1].  The new model is rejected if the random number is greater than the likelihood ratio, and the previous model is duplicated and added to the ``chain."  If the random number is smaller than the likelihood ratio, the model is accepted, added to the chain, and the process is repeated, stepping to another region of parameter space.  Because we are assuming that the RV curve is generated from a star--planet system and can be described by Kepler's Laws, we do not consider the likelihood of models with different parameterizations.    

EXOFAST improves on the MCMC method by implementing Differential Evolution MCMC (DE-MC).  DE-MC is a more elegant approach to the somewhat ``brute force" Monte Carlo method of parameter fitting.  While the regular MCMC method runs a single chain, the DE-MC method will run a number of chains equal to twice the number of free parameters in parallel and calculates the next step in the chain by taking the difference between the parameter values between two random chains.  This approach will dramatically decrease the number of links it takes for the chains to become ``well-mixed" and to sufficiently sample parameter space.  Since it is not practical to use the same step size for many different parameters with different units, EXOFAST adds the additional modification of creating self-adjusting step-sizes by varying each parameter individually until $\Delta\chi^{2} = 1$ and then adding a uniform deviate equal to 1/10 of that step size.  

The parameterization for RV fitting that EXOFAST uses is log\,$P$, log\,$K$, $\sqrt{e}$\,cos\,$\omega_{*}$, $\sqrt{e}$\,sin\,$\omega_{*}$, $T_{C}$, $\gamma$, and $\dot{\gamma}$, where \textit{P} is the period of the orbit, \textit{K} is the RV semi-amplitude (in m/s), \textit{e} is the eccentricity of the orbit (see section \ref{sec:eccentricity}), $\omega_{*}$ is the argument of periastron of the orbit, $T_{C}$ is the time of central transit (not used in RV-only fits), and $\gamma$ is the RV systemic velocity (zero point), in m/s.  The fit will allow us to obtain \textit{K}, \textit{e}, $\omega$, \textit{P}, $T_{P}$, $T_{C}$, as well as derived parameters such as $M_{*}$, $R_{*}$, $m$\,sin\,$i$, and a.  One alteration we made to the standard release of the EXOFAST code was to change the estimate of the stellar mass/radius.  EXOFAST uses the \cite{Torres2009} relation, which was not designed to work for cool stars with mass $M_{*} \,\textless\sim0.6 M_{\odot}$.  We replaced the Torres relation with the \cite{Baraffe1998} models in order to correctly estimate the derived parameters $M_{*}$, $R_{*}$, $m$\,sin\,$i$, and \textit{a}.  \cite{Baraffe1998} agrees reasonably well with \cite{Delfosse2000}, and was chosen over the \cite{Delfosse2000} for consistency, because DEEMA implements the \cite{Baraffe1998} models in order to compute stellar radii (see Paper I).  Since \textit{K}, \textit{e}, and \textit{P} do not depend on the stellar mass/radius relation, they are independent of which model we choose.  

By default, EXOFAST handles uncertainties by scaling the uncertainties by a constant factor.  A jitter term is not included (a systematic such as jitter that is independent of the signal should be added in quadrature), which is a common practice with RV fits.  Although we considered editing the code to include a jitter term, which might improve the quality of the fits, it seemed more useful to carry out analysis of the RV data in the usual way, rather than to account for a systematic error that we are aware of but observational planet hunters would likely not be.  Thus, we analyzed our data in the standard EXOFAST procedure.  

EXOFAST allows the maximum and minimum period for the Lomb-Scargle periodogram to be set, in order to investigate periods that are slightly disfavored in the periodogram.  We used this option to eliminate the rotational period of the star from the results, since this tends to dominate the signal.  EXOFAST returns a large variety of outputs from the fit parameters (see \citealt{Eastman2013}). The parameters used in this work are: $e$, the eccentricity of the orbit; $P$, the period of the orbit, in days; and $a$, the semi-major axis of the orbit, in AU.

\subsection{Period}

\begin{figure}
 \includegraphics[width=84mm]{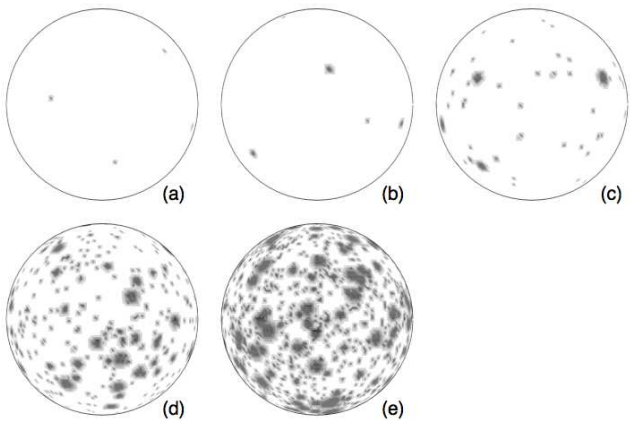} 
\caption{Examples of stellar surface with five different filling factors.  (a): 0.1\%, (b): 0.3\% (approximately the active solar filling factor), (c): 2\%, (d): 10\%, (e): 30\%.}
\label{fig:4ffs_plot}
\end{figure}

We chose a 10 M$_{\oplus}$ planet with a period of 21.2 days orbiting a 0.57\,$M_{\odot}$ M dwarf with a stellar rotation period of 7.7 days. This period corresponds to an orbital radius that would put the planet well inside the classically-defined habitable zone, according to \cite[][see Section \ref{sec:Habitability} for further discussion]{Kopparapu2013}.  We used five filling factors, 0.1\% (Solar), 0.3\% (approximatly the Solar spot filling factor at Solar maximum), 2\%, 10\%, and 30\%. Tests with filling factors above $\sim$30\% resulted in too many of the trials being unable to extract the planetary RV, and thus we were not able to make a meaningful statement about the uncertainties in the derived planetary parameters for these higher filling factors. Example maps of four filling factor cases are shown in Figure ~\ref{fig:4ffs_plot}. For each filling factor, we created 30 random spot maps for each of three different spot size distributions resulting in a total of 90 spot maps per filling factor.  We then ran each spot map through DEEMA, which calculated the RV jitter contribution from the star and added to that the RV created by the planet's orbit at each observational phase.  The result was a curve of radial velocity (including the planetary contribution plus jitter) as a function of time.  

\begin{figure*}
\begin{minipage}{168mm}
 \includegraphics[width=84mm]{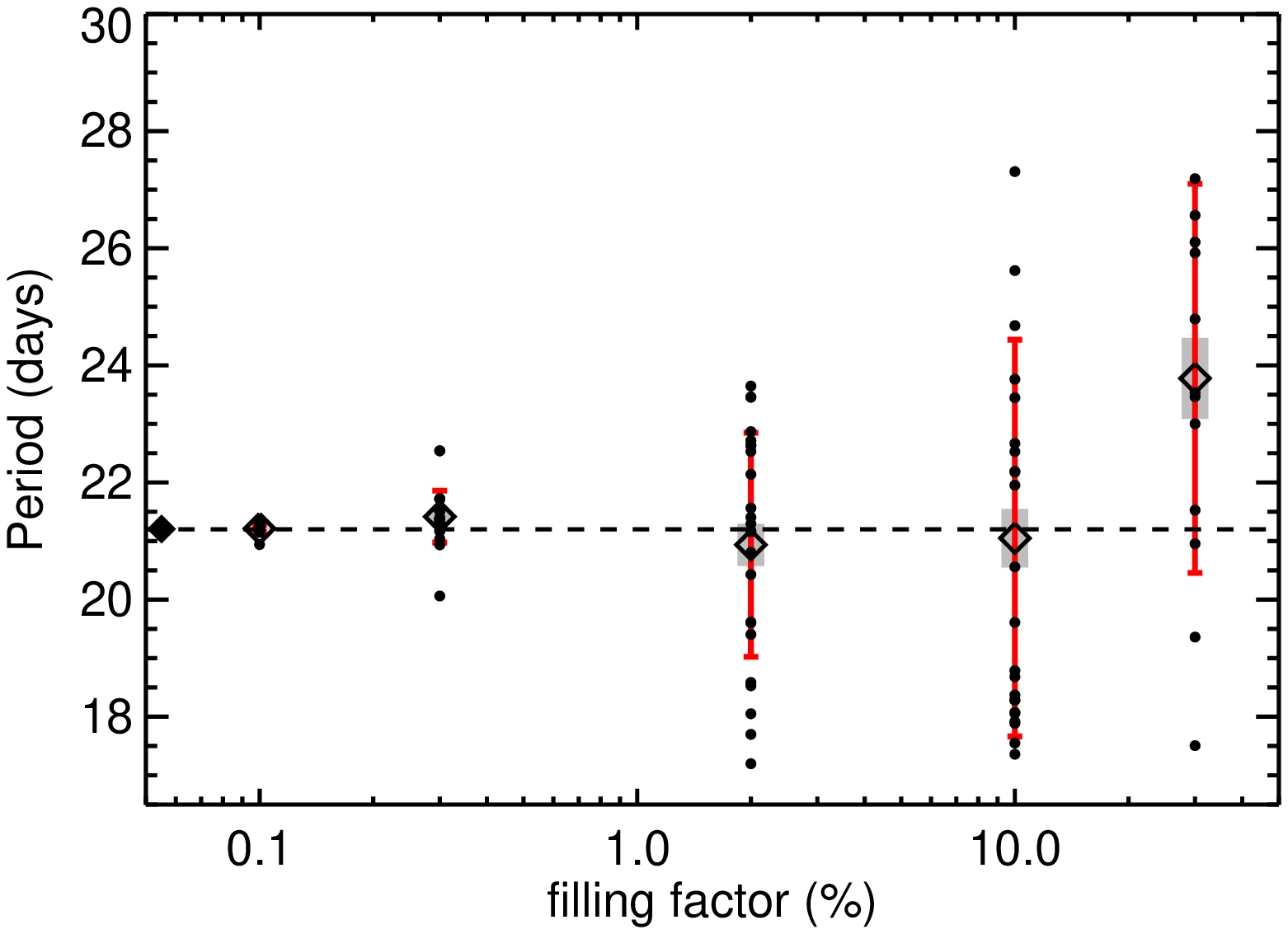} 
  \includegraphics[width=84mm]{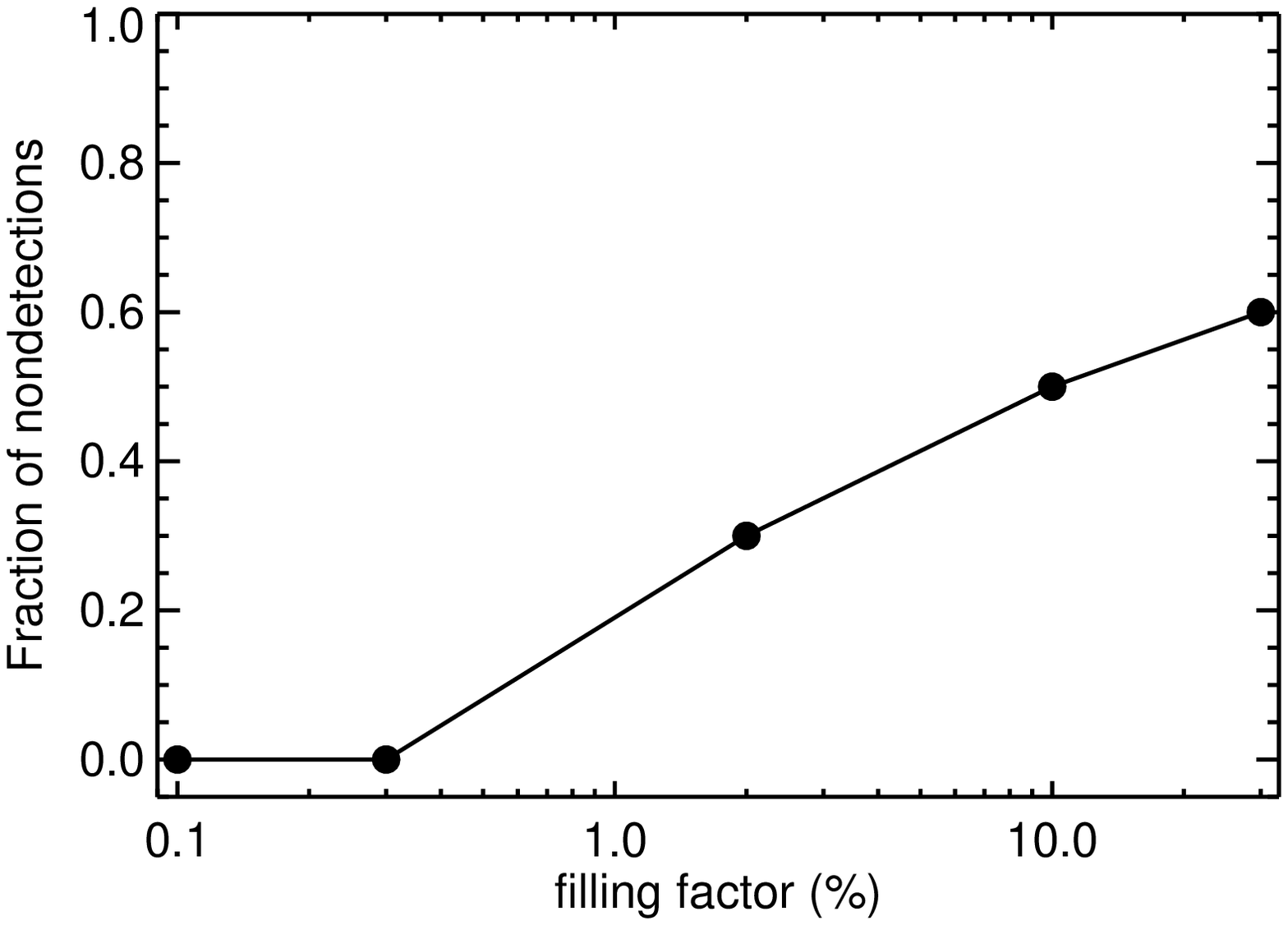} 
\caption{\textit{Left:} Derived periods and uncertainties as a function of spot filling factor. Periods were derived from radial velocity data (spot jitter plus RV from a simulated planetary system) using EXOFAST. Filled circles indicate each individual period derived at a given spot filling factor. Dashed line indicates the period of the planet used to generate the RV cure.  Filled black diamond corresponds to ``zero filling factor" case, i.e., an unspotted star.  Open diamonds correspond to the mean value of all the measured periods at each filling factor. Red error bars show the standard deviation of the measured periods. Shaded gray region indicates the standard error of the mean. {\it Right:} Fraction of nondetections of the orbiting planet for each filling factor.  }
\label{fig:period_plot}
\end{minipage}
\end{figure*}

We took the resulting RV curves and reduced each one with EXOFAST, in an attempt to recover the planetary system parameters.  We used the maximum- and minimum-period input keywords to eliminate the stellar rotation period, which tends to be the strongest periodic signal in the data due to the highly-spotted stellar surfaces.  At each filling factor, we calculated the mean planetary period, the standard deviation, and the standard error of the mean.  We also ran an unspotted surface through the same process for calibration, and this test case is labeled as filling factor of zero.  There is no standard deviation or expectation value for this case since we only used one map.  We found that higher spot filling factors result in higher uncertainty in the final measured planetary period (see Figure \ref{fig:period_plot}). As filling factor increases, the measured values for the period deviate increasingly from that actual value, although the mean value remains close to the actual period until the 30\% filling factor case. Even with small filling factors, e.g., 0.3\%, some of the period measurements are very different from the input period.

As the spot filling factor increased, the percentage of tests where the planetary period was not recovered from the RV curve also increased.  For the Solar filling factors, 0.1\% and 0.3\%, the planetary signature was extracted in all tests.  For filling factors of 2\%, 10\%, and 30\%, the percentages of non-detections were 30\%, 50\%, and 60\%, respectively. Because larger filling factors resulted in many RV curves where the planetary signature is undetectable above the noise from the RV jitter, we focused on filling factors of 30\% and below, although we point out that some active M dwarfs are thought to have filling factors of 50\% or higher. 

\subsection{Eccentricity}
\label{sec:eccentricity}

The choice of eccentricity to use into our planetary system models was rather arbitrary, and we chose an initial eccentricity of zero for two reasons: (1) a circular orbit is the most simple case; and (2) from our investigation of FAP we found that, with a signal well-sampled in phase, planets with higher orbital eccentricities are easier to detect in a LS periodogram. We chose to use a circular orbit which adds the most uncertainty to planet detection using our sampling methods.

\begin{figure}
 \includegraphics[width=84mm]{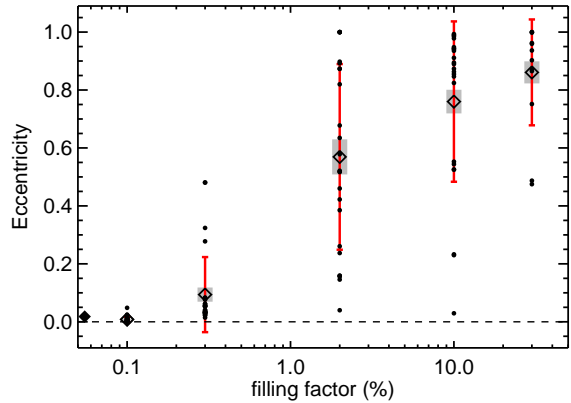} 
\caption{Eccentricity derived from radial velocity curves using EXOFAST, at four spot filling factors. Filled circles indicate each individual eccentricity value derived at corresponding spot filling factor. Dashed line indicates zero eccentricity, the eccentricity value used in generating the simulated RV due to an orbiting planet.  Open diamonds correspond to the mean value of all the measured eccentricities at each filling factor. Red error bars show the standard deviation of the measured eccentricities. Shaded gray region indicates the standard error of the mean. }
\label{fig:ecc_plot}
\end{figure}

We used the same spot maps as in the investigation of period, but instead looked at the derived eccentricity of the planetary orbit from EXOFAST.  The results from the eccentricity study are shown in Figure \ref{fig:ecc_plot}. Similar to Figure \ref{fig:period_plot} we calculated the mean eccentricity, the standard deviation, and the standard error of the mean for all the measurements at each filling factor. When the  filling factor increases, the measured values for the eccentricity deviate substantially from the actual value. For filling factors of 10\% and 30\% basically all the measurements give large eccentricities of $>0.5$, and even a relatively small filling factor of 0.3\% can yield eccentricity measurements as high as 0.5. We acknowledge that at the "zero filling factor" point (filled diamond), i.e. the unspotted star, a greater than zero eccentricity was also measured, likely due to the Lucy-Sweeny Bias (see below). Still, the large measured eccentricities seen with filling factors 0.3\% and larger cannot be explained by the Lucy-Sweeney bias. It is interesting to note that many exoplanets show larger eccentricities than the Solar system planets \citep{Butler2006,Kane2012}, and one possible explanation could be spot activity of the host star. As was shown in Paper I, a spot located at the `right' place can introduce changes in the measured radial velocity curve that can be interpreted as larger eccentricity.

\subsubsection{Eccentricity bias}
There exists a long--understood observational bias against low eccentricities in binary systems due to the data reduction methods of observations of such systems, called the Lucy-Sweeney Bias. Systems with a circular orbit will be found to have a small but nonzero eccentricity due to observational uncertainties and analytical errors.  \cite{Lucy1971} described this bias in binary star systems and how it results from a lack of phase space at exactly $e = 0$, so any uncertainly at all will push the result toward a positive value, since there is no such thing as a negative eccentricity.  \cite{Eastman2013} has dramatically reduced the Lucy-Sweeney bias by allowing a negative eccentricity to be inferred, although they still caution that it is up to the user to inspect the PDF to determine the significance of the calculated eccentricity, and note that orbits with intrinsically small uncertainties will have an over--estimated uncertainty with an uncertainty of roughly 0.007.  Since we use an eccentricity of zero, we expect that the majority of our eccentricities would be slightly overestimated, even without the added RV jitter.

\subsection{Orbital radius and habitability}
 \label{sec:Habitability}
The so-called ``Habitable Zone" around a star is defined as the distance from the star where an orbiting planet could support liquid water on its surface.  It is also sometimes referred to as the ``Goldilocks Zone" since it is ``not too hot, and not too cold, but just right." Though this is a very Earth--centric definition of habitability, it is assumed (and mostly agreed upon) that some form of liquid is necessary for the evolution of life, and liquid water seems to be the most prevalent. \cite{Seager2013} recently redefined the HZ taking into account to the possibility of thick planetary atmospheres trapping heat generated from active planetary interiors, thus extending the HZ farther from the star than was previously expected (and even allowing for the possibility of habitable so-called ``rogue planets"---planets that are not in a stellar orbit.  However, for the purposes of this paper we will adopt the more classic definition of the HZ, and use values from \cite{Kopparapu2013}, which allow for both a ``conservative habitable zone" and an ``optimistic habitable zone".  

The optimistic HZ is bounded by the ``recent Venus limit" on the inner edge and the ``early Mars limit" on the outer edge.  These limits are empirical, based on the assumption that Venus has not hosted liquid water for a least 1 billion years \citep{Solomon1991} 
and the evidence that Mars had liquid water on its surface about 3.8 billion years ago.   Calculating the solar luminosity at these times compared to the present value, a HZ limit can be derived.  More conservative HZ limits rely on the ``water loss"  (inner) and ``maximum greenhouse" (outer) limits, which are simply, according to models, the highest stellar irradiance where all liquid water will not be lost, and the lowest stellar irradiance where the greenhouse effect could still keep a planet's surface sufficiently warm to have liquid water, respectively.

\begin{figure}
 \includegraphics[width=84mm]{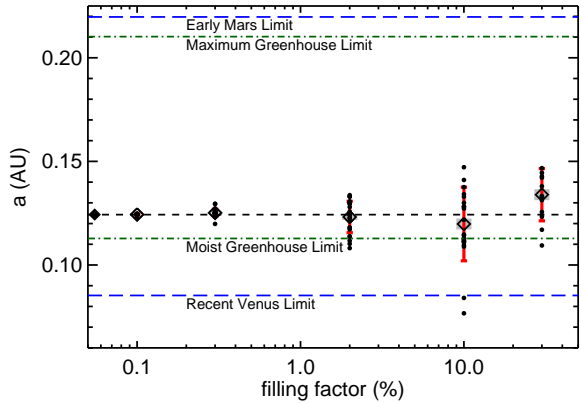} 
\caption{Derived semi-major axis ($a$) of orbit (filled circles) at each of four filling factors, plus ``zero filling factor" which indicates an unspotted surface (filled diamond).  Open diamonds indicate mean values of all measured $a$ values at corresponding filling factor. The standard deviation of the measured values is displayed as red error bars, and the standard error of the mean is indicated by a shaded gray region.  Habitable zone estimates from \citet{Kopparapu2013}  are included.  The conservative HZ (indicated by green dotted lines) is defined by the ''Moist Greenhouse Limit" and the ''Maximum Greenhouse Limit", and the optimistic HZ (indicated by blue dot-dash lines) is bracketed by the ''Recent Venus Limit" and the ''Early Mars Limit".}
\label{fig:HZ_plot}
\end{figure}

Whether or not a planet lies in the HZ depends on the stellar irradiance the planet is receiving, which is highly and sensitively dependent on the star's $T_{eff}$. To calculate an input value using the \cite{Kopparapu2013} HZ equations (also available as an online HZ calculator), we used the M dwarf luminosity from \cite{Casagrande2008}, choosing the M dwarf HIP897 that has an effective temperature of $3786 \pm 126$\,K and a bolometric magnitude of 9.628 mag.  Using this value we calculated $L_{*}/L_{\odot} = 0.011$.  Feeding this value into the \cite{Kopparapu2013} script resulted in the following HZ limits:  conservative,  0.1128 -- 0.2102\,AU, optimistic, 0.0853 -- 0.2197\,AU.  (See Figure \ref{fig:HZ_plot}.)  It would be naive to assume that planetary habitability could be completely determined from total stellar irradiation alone, as there are many other factors that contribute to a planet's habitability.  Since M dwarfs have a very different spectral energy distribution from solar-type stars, planets orbiting M dwarfs will experience a distinct energy environment even at the same stellar irradiation as planets around a G dwarf \citep{Tarter2007}.  The albedo feedback is dependent on the reflectivity of water/ice surfaces, which is wavelength-dependent and will be significantly different for M dwarfs than for the Sun.  Also, the calculations of \cite{Kopparapu2013} do not account for the effects of water clouds, and do not apply to planets that are highly unlike Earth, either in atmospheric composition or surface gravity. These, along with many other factors compel us to present our ``habitable zone" results with some caution.

We used our simulated planet of $P = 21.2$ days around a 0.57\,$M_{\odot}$ M dwarf with eccentricity of zero, which results in an orbital radius of $a = 0.12 $AU.  This is well inside the conservative HZ for this star.  Figure \ref{fig:HZ_plot} shows the measured $a$ for the planet's orbit as a function of spot coverage (filling factor), to demonstrate the uncertainty jitter can contribute to these measurements.  It can be seen that there are cases where the measured value of $a$ falls outside the habitable zone, although the mean and $\sigma$ for most filling factors still lie within this zone. Even at a filling factor as low as 2\%, some of the cases resulted in planetary orbits outside the HZ. With increasing filling factor, the amount of scatter in the derived planetary orbits also increases.  We did not include filling factors above 30\% because at that level of spot coverage too many tests resulted an undetectable planetary signal.  However, for the cases where the planet was still detectable the scatter from the true $a$ value was generally higher, indicating that this increasing trend continues.  

As mentioned previously, high filling factors resulted in high percentages of non-detections for the planetary signal.  In the cases where the planet was detected, we calculated the probability that this HZ planet would be classified as outside the conservative HZ, P(O).  For the Solar filling factors, 0.1\% and 0.3\%, P(O) = 0, meaning all the planet detections were classified as habitable.  The scatter in the measured $a$ values due to jitter, was minimal although still present.  For the higher filling factors, 2\% - 30\%, P(O) = .23.  Thus, in approximately 20\% of cases the HZ planet was classified as outside the HZ.  

All these tests were performed with a planet of mass $10\,M_{\oplus}$.  Lower planetary masses would mean the jitter signal would be higher in comparison to the true RV, and thus we would expect a higher percentage of non detections and more scatter in the derived parameters for a given filling factor.  A more extensive exploration of parameter space including many different eccentricities, planetary masses, stellar inclinations, and spot filling factors is the subject of a future study.   

\section{Conclusions}

In this work we have investigated the effects of activity-induced jitter on planetary system parameters derived from RV fitting.  We modeled the RV jitter using realistic spot models, and improved on previous jitter studies by including active latitude and longitude ranges, evidence of which has been seen in various observations of spot distributions on low-mass stars.  
Since published values of photosphere-to-spot temperature contrasts for M dwarf stars are still highly uncertain, we chose to investigate low, medium, and high contrast spots spanning the range of detected measured temperature contrasts from the literature.  
We quantified the error introduced to derived planetary parameters as a result of RV jitter, focusing on planetary period, eccentricity, and semi-major axis of orbit.  We reached the following conclusions:\\

\textbullet\ Jitter increases with increasing spot contrast at M dwarf photospheric temperatures, which confirms previous studies.  This effect is most prominent at $\sim$5500\,\AA\ where the jitter at low contrast ($\Delta T = 100$) 
is approximately a tenth that of the high contrast ($\Delta T = 900$) 
case.

\textbullet\ RV jitter decreases when moving to longer wavelengths, as has been seen in previous studies. At high contrast, when moving from 3710\,\AA\ to 8210\,\AA\ the jitter is reduced to $\sim$20\% of the jitter at  3710\,\AA. This effect is less extreme at low contrast, but even at $\Delta T = 100$ the jitter at 8210\,\AA\ decreases to half that at 3710\,\AA\ 

\textbullet\ When the spot contrast ratio is kept at a constant $T_{spot}/T_{eff} = 0.90$, the RV jitter is lowest at earliest spectral types (i.e. highest photospheric temperatures).  

\textbullet\ At a given filling factor, the RV jitter from different spot distributions varies significantly.  However, RV jitter increases monotonically with increasing photometric variation, regardless of the spot configuration.  In our tests the relationship was linear. Thus it is more useful to use the spot-induced stellar photometric variation than spot filling factor when trying to predict the RV jitter.  Spot filling factor is not a good predictor of spot-induced jitter. 

\textbullet\  The false alarm probability of planet detection increases with increasing filling factor as well as with increasing contrast between spot and photosphere.  We found that an Earth-sized planet orbiting in the HZ of a $T_{eff} = 3650\,K$ star with a solar filling factor of 1\% could not be detected except in the lowest contrast case.  With a filling factor of 30\% the same star could easily mask planets of 1 and 5 Earth-masses with medium or high contrast, and even at low contrast the 1\,$M_{\oplus}$ is never detectable even up to 100 observations.  In the HZ of a $T_{eff} = 3150\,K$ star, the detection ability was increased.  The Earth-sized planet could be detected in a reasonable number of observations at all contrasts for the 1\ filling factor case, and for 30\% filling factor the Earth-sized planet was detectable at low contrast.  

\textbullet\  Higher orbital eccentricities of the planet result in lower FAPs, meaning planets on a circular orbit are more difficult to detect in the presence of RV noise than those on a more eccentric orbit.  However, the difference in FAP moving from an eccentricity of 0.5 to 0.8 is practically zero in all cases, while the difference between the FAP at 0.0 eccentricity and at 0.3 eccentricity can be quite large (up to 2 orders of magnitude), especially at high filling factors. We note that this will only consistently be true if the RV signals are well-sampled in phase.

\textbullet\  For tests with spot filling factors up to 10\%, the mean and expectation value of the derived planetary period is still close to the actual value.  However, at a 30\% filling factor both the mean and expectation value deviate significantly from the actual period, although the actual period still lies within the standard deviation of all the measured values at this filling factor.   At filling factors higher than $\sim$30\%, a significant number of the tests could not recover the planetary period at all.  

\textbullet\ Due to the Lucy-Sweeny bias, we were not expecting to recover the eccentricity of 0 in the original planetary orbit.  Indeed, even the ``zero filling factor" star, or in other words, the unspotted surface, resulted in an eccentricity of slightly above zero.  However, as the filling factor increases, the measured eccentricity radically increases as well, indicating that with highly spotted stars, one should be wary of a measured eccentricity value. 

\textbullet\  At high filling factors, it is possible for the derived planetary orbits to scatter outside of the conservative HZ, although the mean and expectation values measured at each different filling factor still remain well within these limits.
At filling factors of 2\% - 30\% the probability of classifying the HZ planet as orbiting outside the conservative HZ is 23\%.  
(However, we caution that orbital radius alone is not sufficient to truly determine habitability, and that there are still many factors that are not taken into account when calculating the HZ distance.)
\\

We have shown that stellar RV jitter induced by activity can contribute an important source of noise to planetary measurement that must be accounted for both when searching for and when characterizing potentially habitable planets around active stars.  
Better observational constraints on M dwarf spot parameters--spot sizes, distributions, and temperatures--will help to further constrain the effects of RV jitter on planet detection and characterization.  Stellar photosphere-to-spot temperature contrasts seem to be the most important factor in predicting RV jitter.  Large telescopes with high-resolution spectrographs will permit more Doppler Images of M dwarf spots to be taken in the future, and photometric monitoring from missions such as Kepler and TESS should provide further opportunities to constrain spots using eclipse mapping.  Predicting RV jitter based on photometry alone removes the need to know the actual spot configurations, which is difficult for M dwarf stars due to their low luminosities.  Studies such as \cite{Lanza2011}, \cite{Aigrain2012} \cite{Cegla2014}, and this work have demonstrated the strong correlation between photometric variation and radial velocity jitter, which makes such an approach promising.

\section*{Acknowledgements}
J.M.A. acknowledges support through an NSF Graduate Research Fellowship and the Nordic Research Opportunity award. H.K. acknowledges the support from the European Commission under the Marie Curie IEF Programme in FP7.  

We also wish to thank Uffe Gr\aa e J\o rgensen for comments, Roberto Sanchis-Ojeda for useful discussion on M dwarf spot sizes and temperatures, and R. J. Jackson and R. D. Jeffries for providing insight into the spot behaviors of M dwarfs in NGC 2516.  
Many thanks to Andrew West for proofreading the manuscript and offering advice and constructive criticism, and thank you to the anonymous referee for many useful comments and some thought-provoking questions.

\label{lastpage}

\bibliographystyle{mn2e}
\bibliography{jitter_paper}

\end{document}